\documentclass[aip,jcp,reprint]{revtex4-1}

\draft 

\usepackage{mathtools} 
\usepackage{color}

\pdfoutput=1

\begin{document}


\title{On the use of stereodynamical effects to control cold chemical reactions: the H + D$_{2}\longleftrightarrow$ D + HD case study} 



\author{H. da Silva Jr}
\affiliation{Department of Chemistry and Biochemistry, University of Nevada, Las Vegas, Nevada 89154, USA}

\author{B. K. Kendrick}
\affiliation{Theoretical Division (T-1, MS B221), Los Alamos National Laboratory, Los Alamos, NM 87545, USA}

\author{N. Balakrishnan}
\affiliation{Department of Chemistry and Biochemistry, University of Nevada, Las Vegas, Nevada 89154, USA}


\date{\today}

\begin{abstract}
Quantum calculations are reported for the stereodynamic control of the H + D$_{2}\longleftrightarrow$ D + HD chemical reaction in the energy range of 1\textendash{}50 K. Stereodynamic control is achieved by a formalism similar to that reported by Perreault \textit{et al.} [Nature Chem.  {\bf 10}, 561 (2018)] in recent experimental works in which the alignment of the molecular bond axis relative to the incident relative velocity is controlled by selective preparations of the molecule in a specific or superposition of magnetic projection quantum numbers of the initial molecular rotational level. The approach presented here generalizes the experimental scheme of Perreault \textit{et al.}  and offers additional degree of control through various experimental preparation of the molecular alignment angle. Illustrative results presented for the H + D$_2$ and D + HD reactions show  significant control with the possibility of turning the reaction completely on or off with appropriate stereodynamic preparation of the molecular state. Various scenarios for maximizing and minimizing the reaction outcomes are identified with selective preparation of molecular rotational states.
\end{abstract}


\maketitle 




%
%

%

\section{Introduction}

Stereochemistry \cite{aldegunde2006,aldegunde2011,cernuto2018,ascenzi2019,falcinelli2019,falcinelli2021} along with coherent and optimal control of molecular processes \cite{tannor1986,rice2000,levis2001,shapiro2012} are a manifestation of a long standing goal of controlling chemical reactions. The problem has been addressed independently from the perspective of a variety of research fields; from exploiting ultra\textendash{}fast light\textendash{}induced electron dynamics \textendash{} within an attosecond time scale \cite{cattaneo2001}\textendash{} to an ultra\textendash{}slow sub\textendash{}Kelvin nuclear dynamics, in the context of either cold or ultracold molecular collisions \cite{krems2008,balakrishnan2016}.

In a recent series of seminal experiments, Perreault and co\textendash{}workers have demonstrated a relatively simple setup in which a co\textendash{}expansion of two colliding molecules (HD + D$_{2}$ and HD + H$_{2}$) in the same beam stream enable cold collisions with relative kinetic energies of about 1 Kelvin \cite{perreault2017,perreault2018}. The technique, combined with Stark\textendash{}induced adiabatic Raman passage (SARP) \cite{mukherjee2011} allows selective preparation of the alignment of the diatomic internuclear axis relative to the incident velocity vector. This is made possible by a light\textendash{}assisted spatial polarization of the diatomic angular momentum vector $\mathbf{j}$ combined with selection of specific $\mathbf{j}$\textendash{}projections, say $m$, along a given quantization axis of choice \cite{dong2013,mukherjee2014}. As a consequence, the lower collision energies provided by the co\textendash{}expansion can selectively discard higher and undesired \cite{perreault2017} external collisional orbital angular momentum states, $\ell$, whereas SARP can selectively prepare a specific internal rovibrational state with a given $m$ or superposition of $m$ states. More recently the same methodology was also applied to cold He + HD and He + D$_2$ collisions with the latter revealing strong signatures of an $\ell=2$ partial wave resonance \cite{perreault2019,sarp_hed2}. While these experiments have so far been used to study only inelastic processes, they open up new avenues for possible stereodynamic control of cold  reactive collisions. Theoretical prospects on that matter have been available for a while \cite{aldegunde2005,aldegunde2006,aldegunde2011}, and revisited recently in the context of coherent control \cite{devolder2020}.

It is worthwhile to highlight important aspects of these experiments, beyond requiring a simpler experimental setup compared to current molecule cooling and trapping techniques: (i) it allows controlled low\textendash{}kinetic energy collisions of spatially\textendash{}polarized non\textendash{}polar molecules in a complete field\textendash{}free environment; (ii) the production of colliding partners described by a single internal quantum state eliminates the need of averaging any observable over a large number of initial states; (iii) due to (i) and (ii), the actual interaction potential can be addressed experimentally as it is, within such physical conditions, mainly driven by the anisotropy caused by the target\textendash{}projectile relative orientation; and (iv), unveils marked stereodynamical effects previously constrained to scattering experiments using polar molecules \cite{miranda2011}.

The experiments of Perreault \textit{et al.} have been accompanied by the theoretical works of Croft \textit{et al.} \cite{croft2018,croft2019} on HD + H$_{2}$ and Morita \textit{et al.} \cite{morita2020b,morita2020c} on He + HD collisions. Further, Jambrina \textit{et al.} \cite{2019_prl_jambrina} and Morita \textit{et al.} \cite{morita2020a} have explored the control of isolated partial wave resonances through stereodynamic preparations of HD + H$_2$ and HCl + H$_2$ collisions, respectively, as in the experiments of Perreault \textit{et al.} In a recent work, Jambrina \textit{et al.}  examined stereodynamic control of CO + HD collisions in which the HD molecule is prepared in a stereodynamic state but both molecules undergo a change in rotational angular momentum~\cite{Jambrina_PCCP_2021}.  In this work we extend these ideas to low\textendash energy reactive collisions taking  the prototypical chemical reactions of H + D$_{2}\longleftrightarrow$ D + HD with polarization of the initial rotational state $\mathbf{j}$ of D$_{2}$ and HD. Thus, in what follows, we address questions such as (i) the role of bond axis orientation of D$_{2}$ (HD) relative to the initial relative velocity; (ii) whether it is possible to orient the reactants in a manner to enhance or suppress the production of HD (D$_{2}$); (iii) can we selectively induce the formation of products in a given internal state by controlling the orientation/alignment of the reactants during the collisions? (iv) identify any  propensity rules that accompany these changes. These questions are addressed through a systematic analysis of typical scattering characteristics upon specific choices of so\textendash{}called stereodynamical parameters \textendash{} defined in more details below. Collisions are modeled using a time\textendash{}independent quantum reactive scattering formalism as implemented in the APH3D code~\cite{pack1987,kendrick1999,Kendrick_JPC_2003} that has been used extensively to model atom\textendash{}diatom reactive collisions, combined with the legacy BKMP2 potential energy surface (PES) for the H$_{3}$ system \cite{boothroyd1996}. As we shall see below, we have found strong evidence that supports a nearly\textendash{}total stereodynamic control of a cold chemical reaction, from signifcant enhancement to near complete suppression.

Since the early works of Kuppermann \textit{et al.}~\cite{kuppermann1974} the H + H$_{2}$ exchange reaction and its isotopic variants have been extensively studied in the literature both experimentally and theoretically and we shall focus mainly on aspects related to stereodynamic preparations and collisional outcomes. Interested readers are referred to recent works of Goswami \textit{et al.} \cite{goswami2020} who reported wavepacket calculations of the H + H$_{2}$ exchange reaction; Desrousseaux \textit{et al.} \cite{desrousseaux2018} on rotational excitation in H + HD collisions at temperatures in the range of 10\textendash{}1000 K; recent prediction of a Feshbach resonance on H + HD collisions by Zhou \textit{et al.} \cite{zhou2020}; and quasi\textendash{}classical trajectories simulations by Bossion \textit{et al.} \cite{bossion2018,bossion2020}. Kendrick and co\textendash{}workers have reported extensive studies of geometric phase (GP) effects in cold and ultracold H + H$_2$, H + HD and D + HD collisions with vibrationally excited reactants that revealed a new interference mechanism that controls reactivity in the ultracold regime~\cite{hazra2016,kendrick2016a,kendrick2016b}. Non\textendash{}adiabatic effects in the H + H$_2$ and H + HD reactions with vibrationally excited H$_2$ and HD molecules for vibrational levels $\upsilon=$ 4\textendash{}9 have also been reported by Kendrick \cite{Kendrick_JCP_2018,Kendrick_CP_2018,Kendrick_JPCA_2019}. Several theoretical and combined experiment\textendash{}theory studies on H + D$_{2}$ are also available \cite{wrede1997,banares1998,kendrick1999,fernandez-alonso1999,kendrick2000,kendrick2001,harich2002,kendrick2003,pomerantz2004}.

The paper is organized as follows: Section \ref{sec:theory} provides a brief outline of the theoretical formalism for the stereodynamic preparation. This is followed by results and discussion in Section \ref{sec:results} and a summary and conclusions in Section \ref{sec:summary}.

\section{Theoretical approach \label{sec:theory}}

The methodology used herein is, in many aspects, very similar to that utilized in the previous works of Croft \textit{et al.} \cite{croft2018,croft2019} and Morita \textit{et al.} \cite{morita2020b,morita2020c} except that reactive channels are also taken into consideration as a collision\textendash induced outcome. Thus, upon a H + D$_{2}\left(\upsilon,j\right)$ reactive collision (or likewise D + HD), the probability of detecting a scattered HD$\left(\upsilon^{\prime},j^{\prime}\right)$ product (or D$_{2}$) by a detector placed at $\Omega=(\theta,\phi)$, at a given collision energy $E_{\mathrm{coll}}$, is governed by the state\textendash{}to\textendash{}state scattering amplitude \cite{Kendrick_JPC_2003} $f^{J}_{nn^{\prime}}\left(\theta,E_{\mathrm{coll}}\right)$, \textit{i.e.}

\begin{eqnarray}
\begin{aligned}
f_{nn^{\prime}}^{J}\left(\theta,E_{\mathrm{coll}}\right) & =\frac{2\pi}{k_{n}}\sum_{\ell^{\prime}=\left|J-j^{\prime}\right|}^{J+j^{\prime}}\sum_{\ell=\left|J-j\right|}^{J+j}i^{\left(\ell-\ell^{\prime}+1\right)}\\
 & \times\sqrt{\frac{2\ell+1}{4\pi}}C\left(j\ell J;m,0,m\right)\\
 & \times\sqrt{\frac{2\ell^{\prime}+1}{4\pi}}C\left(j^{\prime}\ell^{\prime}J;m^{\prime},m-m^{\prime},m\right)\\
 & \times d_{0,m-m^{\prime}}^{\ell^{\prime}}\left(\theta\right)\left[\delta_{nn^{\prime}}\delta_{\ell\ell^{\prime}}-S_{n\ell n^{\prime}\ell^{\prime}}^{J}\left(E_{\mathrm{coll}}\right)\right]\label{eq:ScattAmplitudeDef},
\end{aligned}
\end{eqnarray}
where we have multiplied the original scattering amplitude expression of Kendrick, Eq. (22) \cite{Kendrick_JPC_2003}, by $\sqrt{k_{n\prime} / k_{n}}$ and have explicitly expressed the spherical harmonic in terms of the Wigner $d$\textendash{}function. We note that the total scattering amplitude is obtained by summing up all of the contributions from each value of the total angular momentum $\mathbf{J}$:

\begin{equation}
f_{nn^{\prime}}\left(\theta,E_{\mathrm{coll}}\right) = \sum_J\,f_{nn^{\prime}}^{J}\left(\theta,E_{\mathrm{coll}}\right).
\end{equation}

Also, the sums over $\ell$ and $\ell^{\prime}$ effectively average over the azimuthal angle $\phi$ so that Eq. (\ref{eq:ScattAmplitudeDef}) is independent of $\phi$. The initial and final states are denoted by $n=\left(\alpha,\upsilon,j,m\right)$ and $n^{\prime}=\left(\alpha^{\prime}, \upsilon^{\prime},j^{\prime},m^{\prime}\right)$ with $\alpha$ denoting the different atom\textendash{}diatom arrangements, \textit{i.e.} $\alpha=1$ for H + D$_{2}$ and $\alpha=2,3$ for D + HD; $\upsilon jm$ are the vibration, rotation and rotational projection quantum numbers, respectively. Moreover, the vector--coupling coefficients, $C$, are the well known Clebsch--Gordan coefficients and the scattering angle $\theta$ measures the deflection between the asymptotic velocity vectors of H (relative to D$_{2}$) and of D (relative to HD); $S_{n\ell n^{\prime}\ell^{\prime}}^{J}$ is the energy\textendash{}dependent scattering matrix for a given total angular momentum, written in the $\ell$\textendash{}representation $\left(\mathbf{J}=\mathbf{j}+\mathbf{\ell}\right)$, and $k_{n}$ is the magnitude of the wave vector associated to the entrance channel $n$ of the reactants,
\begin{equation}
\frac{\hbar^{2}k_{n}^{2}}{2\mu}= E_{\mathrm{coll}}=E-E_{n}
\end{equation}
where $E_{n}$ denotes the corresponding diatomic eigenvalue, $E$ is the total energy, and $\mu$ is the reduced mass of the reactants.

In the space\textendash{}fixed (SF) reference frame, used for evaluating the scattering matrix, the coordinate origin lies at the center\textendash{}of\textendash{}mass (COM) of the triatomic system. Often, a body\textendash{}fixed (BF) reference frame is convenient for the scattering calculations in which the BF quantization axis lies parallel to the atom\textendash{}diatom relative motion coordinate and coincides with the direction of the initial relative velocity vector. Thus, in this set of coordinates, the rotational projection quantum number $m_{\mathrm{exp}}$ of the molecular state $\left|\alpha\upsilon jm_{\mathrm{exp}}\right\rangle $, prepared experimentally, corresponds to a coherent superposition of all possible pure projections $m$, in $\left\{\left|\alpha\upsilon jm\right\rangle \right\}$, \textit{i.e.}

\begin{equation}
\left|\alpha\upsilon jm_{\mathrm{exp}}\right\rangle =\sum_{m=-j}^{j}d_{mm_{\mathrm{exp}}}^{j}\left(\beta\right)\left|\alpha\upsilon jm\right\rangle \label{eq:DiatomRotStateExpansion}
\end{equation}
where
$\beta$ is the angle between the internuclear axis of D$_{2}$ (or HD) and the incident velocity vector. In the experiments of Perreault \textit{et al}. $\beta$ denotes the orientation between the polarization lasers and the molecular beam axis; where choices of $\beta$ correspond to either a parallel orientation $\left(\beta=0\right)$, known as a H\textendash{}SARP preparation, or a perpendicular orientation $\left(\beta=90^{\circ}\right)$, a V\textendash{}SARP preparation. Here, H\textendash{} and V\textendash{} denote horizontal and vertical polarizations, respectively. A cross polarization state, referred to as X\textendash{}SARP, corresponding to a superposition of $m=\pm 1$, was also adopted in their recent work on He + HD and He + D$_2$ collisions \cite{perreault2019,sarp_hed2}. For exploring possible control scenarios, several values of $\beta$ within $\left[0,90^{\circ}\right]$ are considered below.

As discussed by Croft and Balakrishnan \cite{croft2019}, since the experimental results correspond to an integral over the azimuthal angle $\phi$, one can express the differential cross sections (DCS) for any SARP\textendash{}prepared state as a sum of contributions from each $m$\textendash{}state weighted by $\left|d_{mm_{\mathrm{exp}}}^{j}\left(\beta\right)\right|^{2}$. For the state\textendash{}to\textendash{}state DCS, this yields

\begin{equation}
\frac{d\sigma_{\widetilde{n}n^{\prime}}\left(\theta,E_{\mathrm{coll}}\right)}{d\Omega}=\sum_{m=-j}^{j}\left|f_{nn^{\prime}}\left(\theta,E_{\mathrm{coll}}\right)\right|^{2}\left|d_{mm_{\mathrm{exp}}}^{j}\right|^{2}\label{eq:DiffCrossSectionDef}
\end{equation}
where, $\widetilde{n}=\left(\alpha,\upsilon,j,m_{\mathrm{exp}}\right)$. Likewise, for the integral cross section (ICS),
\begin{equation}
\sigma_{\widetilde{n}n^{\prime}}\left(E_{\mathrm{coll}}\right)=2\pi\int_{0}^{\pi}\frac{d\sigma_{\widetilde{n}n^{\prime}}\left(\theta,E_{\mathrm{coll}}\right)}{d\Omega}\sin\left(\theta\right)d\theta. \label{eq:IntCrossSectionDef}
\end{equation}

\begin{figure}
\includegraphics[scale=0.4]{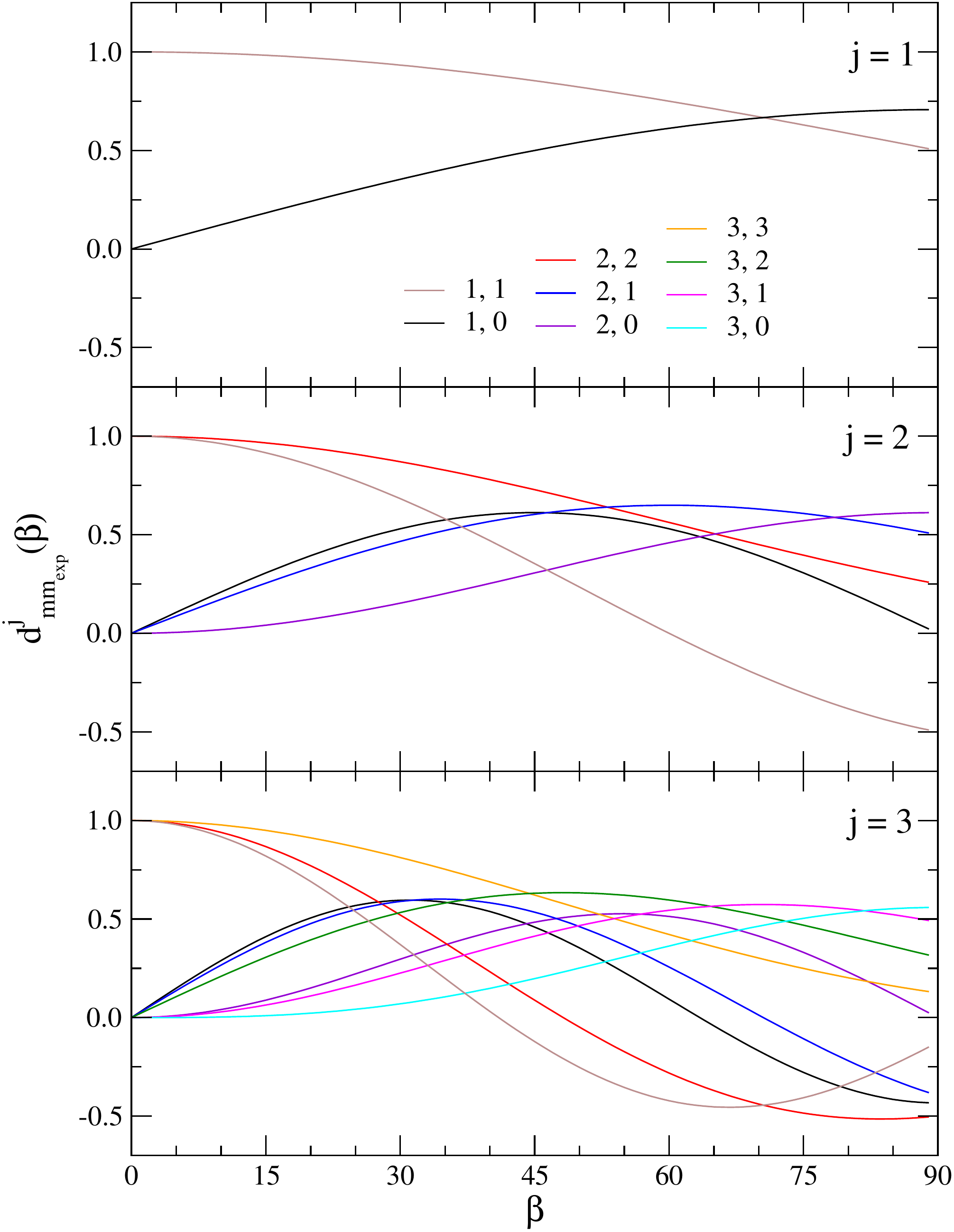}
\caption{\label{fig:WignerD}Wigner\textquoteright s reduced rotation matrix, $d_{mm_{\mathrm{exp}}}^{j}\left(\beta\right)$, as function of $\beta\in\left[0,90^{\circ}\right]$, in degree, for a few cases of $m$, $m_{\mathrm{exp}}$ (solid lines) and $j=1,2,3$ (panels). See the main text for the definition of $m_{\mathrm{exp}}$.}
\end{figure}

Figure (\ref{fig:WignerD}) shows the $\beta$\textendash{}dependence of the Wigner's reduced rotation matrices for the hypothetical cases of $j=1,2,3$ along with their pair of $\left(m,m_{\mathrm{exp}}\right)$ values (negative values are not shown). Upon a simple inspection of Fig. (\ref{fig:WignerD}), the following aspects are apparent:
(i) For $\beta=0$, $d_{mm_{\mathrm{exp}}}^{j}=0$ whenever off\textendash diagonal elements are concerned, \textit{i.e.} $m_{\mathrm{exp}}\neq m$;
(ii) similarly, $d_{mm_{\mathrm{exp}}}^{j}=1$, for diagonal terms at $\beta=0$ and, as a consequence, parallel orientations of the reactants are likely to be composed by terms driven by $m_{\mathrm{exp}}=m$,  \textit{i.e.} $\left|\alpha\upsilon jm_{\mathrm{exp}}\right\rangle =\left|\alpha\upsilon jm\right\rangle $;
(iii) in the particular case of $m_{\mathrm{exp}}=0$ and $\beta=0$, due to (ii), the expansion of Eq. (\ref{eq:DiatomRotStateExpansion}) collapses to a single and isotropic  term in which $m_{\mathrm{exp}}=m=0$;
(iv) $d_{00}^{0}\left(\beta\right)=1$ for any value of $\beta$ \textendash{} not shown in Fig. (\ref{fig:WignerD}) \textendash{} and, thus, as expected, processes associated to rotationless reactants are invariant upon changes of $\beta$.

For the sake of clarity, it is worthwhile to illustrate at least a few possible linear combinations shown by Eq. (\ref{eq:DiatomRotStateExpansion}). Consider a reactant prepared in the rotational state $j=2$ of a given vibrational manifold $\upsilon$ with a projection $m_{\mathrm{exp}}=0$, and subjected to two orientations of its internuclear axis relative to the incoming relative velocity vector: $\beta=0$ and $\beta=45^\circ$. The first case, $\beta=0$, has been addressed above and the resultant state is simply $\left|\alpha\upsilon j;m_{\mathrm{exp}}=0\right\rangle =\left|\alpha\upsilon j;m=0\right\rangle$. Often, the $m=0$ contribution in the ICS (DCS), for uncontrolled collisions, is dominant but suppressed by the typical average made up of all $2j+1$ initial $m$\textendash terms. Thus, we may expect that a fixed orientation of $\beta=0$ for a $m_{\mathrm{exp}}=0$ preparation, in which only $m=0$ contributes, to enhance substantially the collision\textendash{}induced state\textendash{}to\textendash{}state branching ratio of interest. If, instead, a $\beta=45^\circ$ alignment is made, the same $m_{\mathrm{exp}}=0$ preparation is expected to be composed by a broad mixture of all $m$\textendash{}terms, namely:
\begin{eqnarray}
\begin{aligned}
\left|\alpha\upsilon j;m_{\mathrm{exp}}=0\right\rangle & =0.306\left|\alpha\upsilon j;m=-2\right\rangle\\
 & -0.612\left|\alpha\upsilon j;m=-1\right\rangle\\
 & +0.250\left|\alpha\upsilon j;m=0\right\rangle\\
 & +0.612\left|\alpha\upsilon j;m=+1\right\rangle\\
 & +0.306\left|\alpha\upsilon j;m=+2\right\rangle,
\end{aligned}
\label{eqn:sarp_beta=45}
\end{eqnarray}
which is now mainly dominated by contributions from $m=\pm1$ and $m=\pm2$ with a minor contribution from $m=0$. Note that in Eq.(\ref{eqn:sarp_beta=45}) (and in similar expressions below) we have used numerical values of $d_{mm_{\mathrm{exp}}}^{j}(\beta)$ matrix elements corresponding to the chosen $\beta$ values. If a $m_{\mathrm{exp}}=2$ preparation is considered at the same $\beta=45^\circ$ alignment, the linear combination of $m$\textendash{}terms yields
\begin{eqnarray}
\begin{aligned}
\left|\alpha\upsilon j;m_{\mathrm{exp}}=2\right\rangle & =0.021\left|\alpha\upsilon j;m=-2\right\rangle\\
 & -0.103\left|\alpha\upsilon j;m=-1\right\rangle\\
 & +0.306\left|\alpha\upsilon j;m=0\right\rangle\\
 & -0.603\left|\alpha\upsilon j;m=+1\right\rangle\\
 & +0.728\left|\alpha\upsilon j;m=+2\right\rangle.
\end{aligned}
\end{eqnarray}

In this case, the prepared state is mostly composed of $m=+2$ and $m=+1$, and the $m=0$ term. The preparation for $j=2$, $m_{\mathrm{exp}}=0$ at the two typical orientations of $\beta=0$ and $\beta=90^{\circ}$ (V\textendash{}SARP and H\textendash{}SARP) has been discussed in great detail by Perreault \textit{et al.} \cite{perreault2017,perreault2018}.

The scattering amplitudes in Eq. (\ref{eq:ScattAmplitudeDef}) are computed using a numerically exact time\textendash{}independent quantum reactive scattering formalism based on hyperspherical coordinates as implemented in the APH3D code~\cite{pack1987,kendrick1999,kendrick2000,kendrick2001,kendrick2003}. Details of the computational formalism have been discussed in great detail in prior works of Kendrick~\cite{kendrick1999,kendrick2000,kendrick2001,kendrick2003} and in  recent works of Kendrick \textit{et al.}~\cite{kendrick2016b} on the D + HD and  H + D$_{2}$ reactions. We shall remark only on some key details here. The hyper\textendash{}radius $\rho$ describing the radial motion of the triatomic system is divided into an inner region where all three atoms are in close proximity and an outer region where the different atom\textendash{}diatom arrangement channels are uncoupled. In the inner region the three\textendash{}body Hamiltonian is expressed in the adiabatically adjusting principal axis hyperspherical (APH) method of Pack and Parker \cite{pack1987}. Smith\textendash{}Johnson hyperspherical coordinates \cite{smith1962b,kendrick1999} are used in this region where collision\textendash{}induced re\textendash{}arrangements are more likely to occur, and Delves hyperspherical coordinates \cite{delves1958,delves1960,parker2002} are used in the outer region. An in\textendash{}depth discussion of the hyperspherical coupled channel equations and numerical strategies to solve them are available elsewhere \cite{kendrick1999,kendrick2000,kendrick2001,kendrick2003}. Briefly, the six\textendash{}dimensional three\textendash{}body problem is reduced to a set of one\textendash{}dimensional coupled equations along the scattering coordinate $\rho$, that is discretized in a grid of $N$ sectors. The eigenvalue problem associated to the remaining five (internal) degrees of freedom is solved adiabatically, at each sector, yielding an effective set of coupled potentials driving the relative motion along $\rho$. The five\textendash{}dimensional (5D) eigenvalue problem is solved in the APH region by means of an implicitly restarted Lanczos method for sparse matrices as implemented in the PARPACK library \cite{sorensen1992,maschhoff1996}. The corresponding eigenvalues within the Delves region are evaluated using a 1D Numerov propagator \cite{johnson1977}. Once the eigenvalue problem is solved in both regions  for all sectors and the sector\textendash{}to\textendash{}sector overlap and transformation matrices evaluated, the resulting set of radial coupled equations are solved using Johnson's log\textendash{}derivative method \cite{johnson1973}, first from $\rho_{\mathrm{min}}$ to $\rho_{\mathrm{match}}$. At $\rho_{\mathrm{match}}$ the numerical solutions, from the outermost sector of the APH region, are projected onto solutions at the innermost sector of the Delves region. The propagation is continued from $\rho_{\mathrm{match}}$ to $\rho_{\mathrm{max}}$, a sufficiently large value of $\rho$ where the interaction potential is negligible. At  $\rho_{\mathrm{max}}$ all channels (from all arrangements) are numerically decoupled, and scattering boundary conditions are applied to the log\textendash{}derivative matrix in order to evaluate the scattering S\textendash{}matrix. This involves projecting the log\textendash{}derivative matrix onto solutions associated with each asymptotic diatomic state, in appropriate Jacobi coordinates, yielding the $S_{n\ell n^{\prime}\ell^{\prime}}^{J}$ matrix \cite{pack1987}. The procedure is repeated independently for each value of the total angular momentum quantum number $J$, $J$\textendash{}parity $P$ (good quantum numbers in the absence of external forces) and, each $j$\textendash{}parity $p$ for the D$_{2}$ molecule.
It should be noted that the basis sets for both APH and Delves regions are independent of collision energy and, therefore, evaluated only once for each $\left(p,P,J\right)$ sets. Also, both the APH and Delves solutions are properly symmetrized with respect to the identical particle permutation symmetry (\textit{e.g.}, the two $D$ nuclei in HD$_2$).

As both H + D$_{2}$ and D + HD chemical reactions (as well as other isotopic combinations) have been modeled with the APH3D code in the past \cite{kendrick2001}, the optimal set of parameters required to extract fairly converged ($< 5\%$) scattering characteristics are well established. In this work we follow closely the parameterization used previously by our group \cite{kendrick2000,kendrick2001,kendrick2003,kendrick2016a,kendrick2016b}. Thus, the APH region ranges from $\rho_{\mathrm{min}}=1.6$ $a_{0}$ to $\rho_{\mathrm{match}}=7.8$ $a_{0}$ with 64 sectors varying logarithmically with a step size of $\triangle\rho_{\mathrm{aph}}=0.025$ $a_{0}$. The Delves region, starting at $\rho_{\mathrm{match}}$, extends to  $\rho_{\mathrm{max}}=78$ $a_{0}$ with 350 sectors varying linearly with a step size $\triangle\rho_{\mathrm{delves}}=0.2$ $a_{0}$. Reactions such as H + D$_{2}$ are known to require a few dozens of partial waves whenever collision energies exceed 1 eV ~\cite{kendrick2001,kendrick2003}. Here however, we shall simulate the reaction at low and moderate collision energies, between 1 K (about 10$^{-5}$ eV), a typical value in the experiments of Perreault and co\textendash{}workers, and 50 K (about 10$^{-3}$ eV). Thus, to reduce computational efforts associated with nonzero initial rotational levels, we limit our calculations to sets of $J_{\mathrm{max}}=$ 4, 7 and 9 depending on the entrance channel taken into consideration. Numerical convergence is estimated within $1\%$ and, as we shall see below, most of the interesting physics is captured at resonant energies where the discrete contribution of specific (and often lower) values of $J$ is more relevant.

In order to compute the APH basis set, the 64 sectors between $\rho_{\mathrm{min}}$ and $\rho_{\mathrm{match}}$ are further subdivided into five groups each of which with increasing maximum values of $l$ and $m$ (not to be confused with the orbital angular momentum and its projection quantum numbers), two parameters that define the primitive basis sets used in the APH region \cite{kendrick1999}: (i) $l_{\mathrm{max}}=103$, $m_{\mathrm{max}}=190$ for $\rho=$ 1.6\textendash{}2.9; (ii) $l_{\mathrm{max}}=115$, $m_{\mathrm{max}}=214$ for $\rho=$ 2.9\textendash{}3.7; (iii) $l_{\mathrm{max}}=123$, $m_{\mathrm{max}}=232$ for $\rho=$ 3.7\textendash{}4.6; (iv) $l_{\mathrm{max}}=135$, $m_{\mathrm{max}}=250$ for $\rho=$ 4.6\textendash{}5.8; and, (v) $l_{\mathrm{max}}=143$, $m_{\mathrm{max}}=274$ for $\rho=$ 5.8\textendash{}7.8. Finally, 350 $\left(J=0\right)$, 650 $\left(J=1\right)$, 950 $\left(J=2\right)$, 1250 $\left(J=3\right)$, 1550 $\left(J=4\right)$ and 1850 $\left(J=5\right)$ channels are propagated for batches of 50 collision energies at each of the entrance channels described below and every parity of $J$ and $j$. For the $J>5$ case a parity--adapted basis--set is utilized: 2170 $\left(J=6,7\right)$, 2700 $\left(J=8,9\right)$ channels for even $P$; 1860 $\left(J=6\right)$, 2400 $\left(J=7,8\right)$ and 3000 $\left(J=9\right)$ channels for odd $P$.

\section{Results \label{sec:results}}

In what follows the collision\textendash{}induced exothermic reaction $\left(\alpha\neq\alpha^{\prime}\right)$ and inelastic de\textendash{}excitation $\left(\alpha=\alpha^{\prime}\right)$ are studied for the reactants in the initial rovibrational state $\upsilon=4,j=$ 1, 2 $(J_{\mathrm{max}}=4)$, 3 $(J_{\mathrm{max}}=7)$ and 4 $(J_{\mathrm{max}}=9)$; whereas the internal states of the products are chosen to be either $\upsilon^{\prime}=\upsilon$ or $\upsilon^{\prime}=\upsilon-1$ as well as $j^{\prime}=$ 0\textendash{}3. Choices of initial and final states are somewhat arbitrary and intended as representative cases upon which an actual experiment could be designed. However, it should be noted that, as pointed out by Hazra \textit{et al.} \cite{hazra2016}, the effective reaction pathway along the set of adiabatic coupled potentials, for hydrogen exchange reactions, becomes barrierless for the vibrational manifolds above $\upsilon=3$, which is the primary motivation underlying the choice of $\upsilon=4$ herein, since a natural boost in terms of reactivity is expected for low energy collisions \cite{kendrick2016b}.

From an experimental perspective there is a growing effort towards addressing even higher vibrational states \cite{perreault2016,mukherjee2017} in order to explore the inherent delocalization of such weakly\textendash{}bound systems \cite{perreault2020}. Indeed, Perreault \textit{et al.} have recently reported complete transfer of population of H$_2(v=0,j=0$) to H$_2(v=7,j=0)$ via the SARP method opening pathways for collisional experiments involving highly vibrationally excited H$_2$. From a theoretical standpoint however, the description of highly excited vibrational levels during collisions would likely require a treatment which takes into consideration either (i) couplings to higher electronic states or (ii) geometric phase effects  whenever the nuclear motion encircles the conical intersection between the two lowest electronic states. In particular, Kendrick and co\textendash{}workers have discussed both the sources of geometrical phase effects and their consequences on the title reactions for reactants prepared at $\upsilon=4,j=0$ and below \cite{kendrick2000,kendrick2016a,kendrick2016b}. The effects are found to play a negligible role for exchange reactions at collision energies above 1 K but to have a somewhat stronger influence on the inelastic outcome  in the cold and ultracold regimes.

Previous studies of Croft \textit{et al.}~\cite{croft2018,croft2019} and Morita \textit{et al.} ~\cite{morita2020b,morita2020c} have shown that collision\textendash{}induced resonances are more sensitive to the relative orientation of the colliding partners. Therefore, we have first scanned collision energies for the various entrance channels mentioned above to survey for any resonances. To our knowledge resonances in the domain up to 50 K above the $\upsilon=4,j=1-4$ thresholds of D$_{2}$ and HD have not yet been reported. We have identified several instances of these and, given the energy regimes involved, are likely to be shape resonances. A detailed characterization in terms of the angular momentum partial waves, width, lifetime and bound state structure of the  effective potential well will be addressed elsewhere. Here, we focus on the sensitivity of these resonances to stereodynamic preparation, in particular, to choices of both $m_{\mathrm{exp}}$ and $\beta$ in a manner to simulate physical conditions that resemble typical experimental conditions of Perreault \textit{et al}. Despite strict choices of $m_{\mathrm{exp}}=0$, $\beta=0, \beta=90^{\circ}$ and $E_{\mathrm{coll}}\approx$ 1--5 K in the experiments reported so far, we take some (theoretical) freedom to explore other possibilities.

\subsection{Integral cross sections}

\begin{figure}
\includegraphics[scale=0.42]{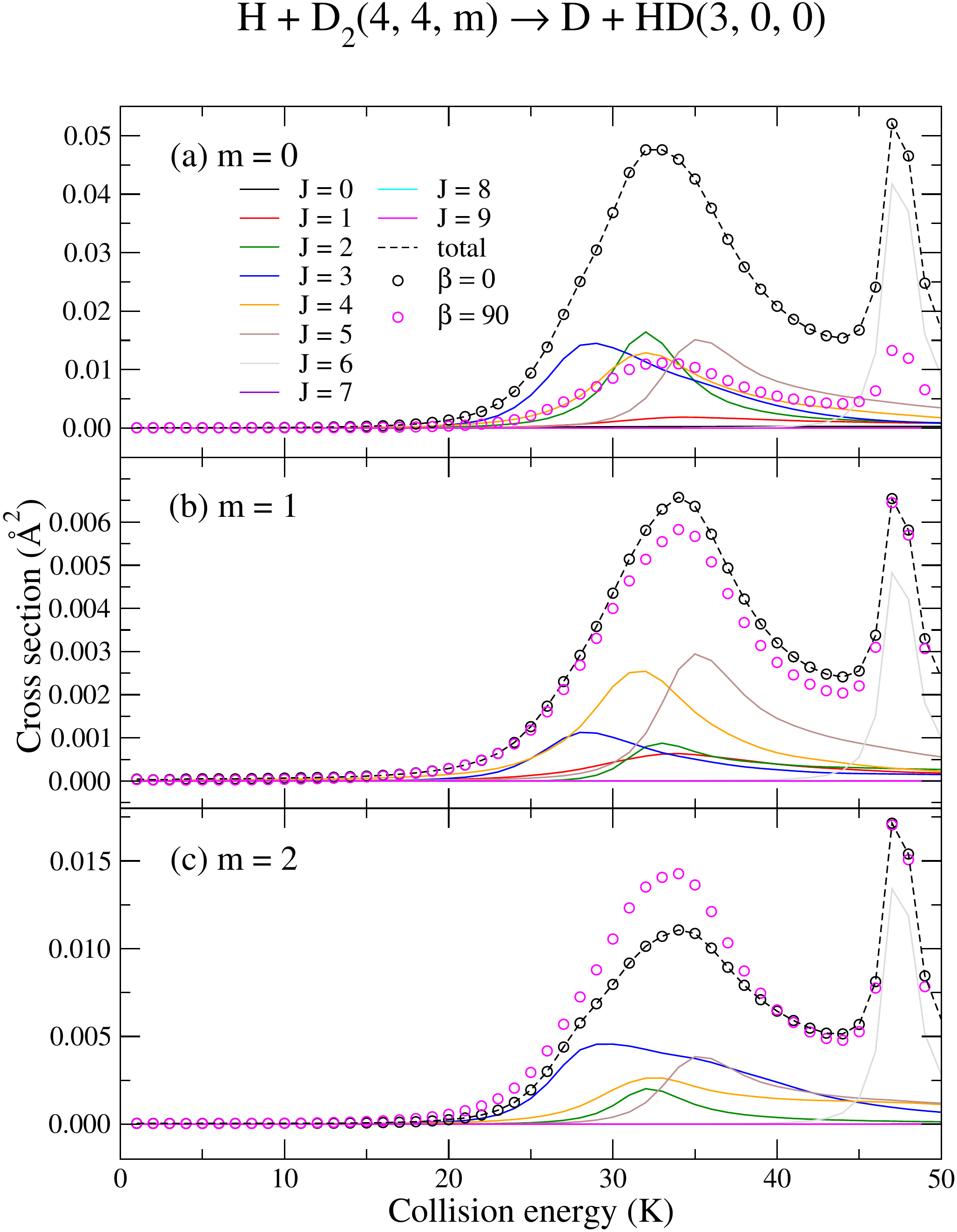}

\caption{\label{fig:ICSvsCollEnergy44-30First} Panels (a)\textendash{}(c): partial ($J=$ 0\textendash{}9) and total $m$\textendash{}dependent ICS as functions of the collision energy for the H + D$_{2}\left(\upsilon=4,j=4,m=0,1,2\right)\protect\longrightarrow$ D + HD$\left(\upsilon^{\prime}=3,j^{\prime}=0,m^{\prime}=0\right)$ chemical reaction; $\beta=0$ (black circles) and $\beta=90^{\circ}$ (magenta circles) are used for the corresponding $m_{\mathrm{exp}}=m$ cases.}
\end{figure}

\begin{figure}
\includegraphics[scale=0.42]{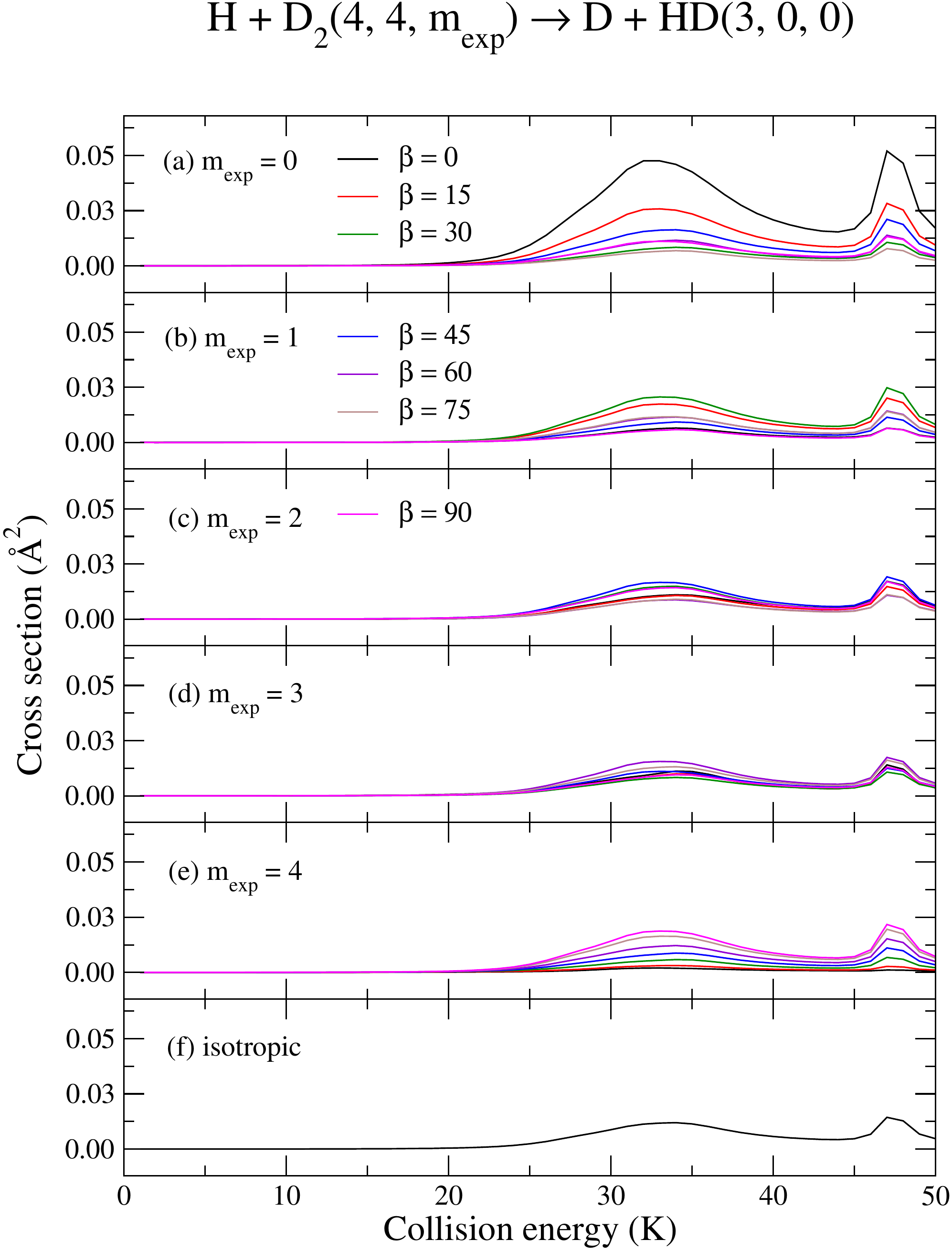}

\caption{\label{fig:ICSvsCollEnergy44-30Second} Panels (a)\textendash{}(e): $m_{\mathrm{exp}}$\textendash{}dependent ICS as functions of the collision energy for the same reaction of Fig. (\ref{fig:ICSvsCollEnergy44-30First});  $\beta=0$ (black curve), $\beta=15^{\circ}$ (red curve), $\beta=30^{\circ}$ (green curve), $\beta=45^{\circ}$ (blue curve), $\beta=60^{\circ}$ (violet curve), $\beta=75^{\circ}$ (brown curve) and $\beta=90^{\circ}$ (magenta curve). Panel (f): total ICS as a function of the collision energy averaged over all $2j+1$ initial $m$\textendash{}values and summed over all $2j^{\prime}+1$ final $m^{\prime}$\textendash{}values.}
\end{figure}

Panels (a)\textendash{}(c) of Fig. (\ref{fig:ICSvsCollEnergy44-30First}) present the $m$\textendash{}dependent ICS as functions of the collision energy for the H + D$_{2}\left(\upsilon=4,j=4,m\right)\longrightarrow$ D + HD$\left(\upsilon^{\prime}=3,j^{\prime}=0,m^{\prime}=0\right)$ reaction, where solid lines are used for each $J$\textendash{}contribution of a given $m$\textendash{}value ($m>2$ and negative values are not shown for the sake of clarity). Needless to say, both resonant features depicted at about 32 K and 47 K are mostly dominated by the $m=0$ component (panel a) of mid--range partial waves, $J=$ 2\textendash{}6, whereas the $J=6$ term alone is responsible for the resonance at 47 K. Also, it is worthwhile to notice that $J=$ 7\textendash{}9 provide just a minor correction which suggests that numerical convergence is attained at $J_{\mathrm{max}}=6$.

Yet panels (a)\textendash{}(c) offer a first glimpse on the stereodynamical effects underlying the reaction where hollow symbols denote the ICS for which the initial rotational state of D$_{2}$, $j=4$, is prepared in a specific projection $m_{\mathrm{exp}}$, hence a weighted sum of various $2j+1$ $m$\textendash{}terms in $-4\leq m\leq4$ contribute, as shown in Eq. (\ref{eq:DiffCrossSectionDef}). Results are shown for internuclear axis orientation $\beta=0$ (black circles) and $\beta=90^\circ$ (magenta circles) with respect to the incoming relative velocity vector. Regardless of the choice of $m_{\mathrm{exp}}$, the  orientation of the internuclear axis alone is capable of modulating the ICS and, as we shall see below, the appropriate choice of a given $m_{\mathrm{exp}}$ and $\beta$ combined, can effectively enhance (suppress) the chemical reaction at the resonant energy. The $m_{\mathrm{exp}}=0$ and $\beta=0$ (panel a) represents the special case, mentioned in the previous section, in which all off\textendash{}diagonal elements of $d_{mm_{\textrm{exp}}}^{j}$ are zero, whenever $m\neq0$, \textit{i.e.} $d_{m0}^{4}\left(0\right)=0$. Therefore, the ICS expansion is reduced to a sole term associated to $m=0$, where $d_{00}^{4}\left(0\right)=1$. In order to better visualize why the same $m_{\mathrm{exp}}=0$ preparation yields a much smaller cross section at $\beta=90^\circ$, panels (a)\textendash{}(b), we invoke Eq. (\ref{eq:DiatomRotStateExpansion}) again to inspect the actual linear expansion for the prepared state:

\begin{eqnarray}
\begin{aligned}
\left|\alpha\upsilon j;m_{\mathrm{exp}}=0\right\rangle & =0.523\left|\alpha\upsilon j;m=-4\right\rangle \\
 & -0.395\left|\alpha\upsilon j;m=-2\right\rangle \\
 & +0.375\left|\alpha\upsilon j;m=0\right\rangle  \\
 & -0.395\left|\alpha\upsilon j;m=+2\right\rangle \\
 & +0.523\left|\alpha\upsilon j;m=+4\right\rangle.
\end{aligned}
\end{eqnarray}

Where, we note that $d_{\pm3,0}^{4}\left(90^{\circ}\right)=d_{\pm1,0}^{4}\left(90^{\circ}\right)=0$, and as a result, $m=\pm1$ and $m=\pm3$ do not contribute to the prepared state. Thus, inspecting panel (a) of Fig. (\ref{fig:ICSvsCollEnergy44-30First}) it becomes clear how a perpendicular orientation $\left(\beta=90^{\circ}\right)$ composed of a broader mixture of $m$ contributions that favor $m=\pm4$ and $m=\pm2$ yields a smaller ICS for $m_{\mathrm{exp}}=0$. The prepared state has only a smaller contribution from $m=0$ that features the largest ICS. A similar analysis applies to results in panels (b) and (c) of Fig. (\ref{fig:ICSvsCollEnergy44-30First}) for $m_{\mathrm{exp}}=$ 1 and 2. In the particular case of panel (c), the relatively smaller $m=2$ component is now mostly favored by the mixture created with a $\beta=90^{\circ}$ alignment (comparing the magenta circles and the dashed black curve of panel c). As we shall see below, higher $m_{\mathrm{exp}}$ values are systematically favored by higher $\beta$\textendash{}values.

For convenience, in the remainder of this paper,  results presented will correspond to a sum over all $J$\textendash{}values. However, whenever appropriate, we shall address the various partial wave contributions in terms of the orbital angular momentum quantum number $\ell$ with $\left|J-j\right|\leq\ell\leq\left(J+j\right)$.

Panels (a)\textendash{}(e) of Fig (\ref{fig:ICSvsCollEnergy44-30Second}) present the ICSs for each case of $m_{\mathrm{exp}}\leq j$ and $\beta=0$\textendash{}$90^\circ$, whereas panel (f) shows the so\textendash{}called isotropic ICS, usually evaluated in ordinary studies of reactive scattering and represents a collision with no control imposed upon the rotational state $\mathbf{j}$. The isotropic case corresponds, therefore, to an ICS summed over all final $m^{\prime}$ (in this particular example happens to be only $m^{\prime}=0$ due to $j^{\prime}=0$) and averaged over all $2j+1$ initial $m$\textendash{}values. The peak predicted at about 32 K is mostly composed of partial waves $\ell=$ 0--7 whereas the one at 47 K is mainly composed by even terms, $\ell=$ 2, 4, 6 and 8. All contributions from partial waves higher than $\ell=9$ are negligible.

Inspection of panels (a)\textendash{}(e), unveils a propensity in which reactivity associated to lower projections of $j$, $m_{\mathrm{exp}}\approx$ 0\textendash{}1, may be significantly enhanced by near parallel orientations of the reactants, say $\beta\leq15^{\circ}$, whereas higher projection values, $m_{\mathrm{exp}}>2$, are boosted by higher orientation angles. Almost full suppression of the chemical reaction is attained by preparing the reactants in an appropriate superposition of rotational projection states corresponding to specific orientation of its internuclear axis (with respect the incident relative velocity vector). Typical cases are depicted in Fig (\ref{fig:ICSvsCollEnergy44-30Second}), namely: $m_{\mathrm{exp}}=0$, $\beta=75^{\circ}$ (panel a, brown curve); $m_{\mathrm{exp}}=1$, $\beta=90^{\circ}$ (panel b, magenta curve); and, $m_{\mathrm{exp}}=4$, $\beta=0$ (panel e, black curve).

\begin{figure}
\includegraphics[scale=0.43]{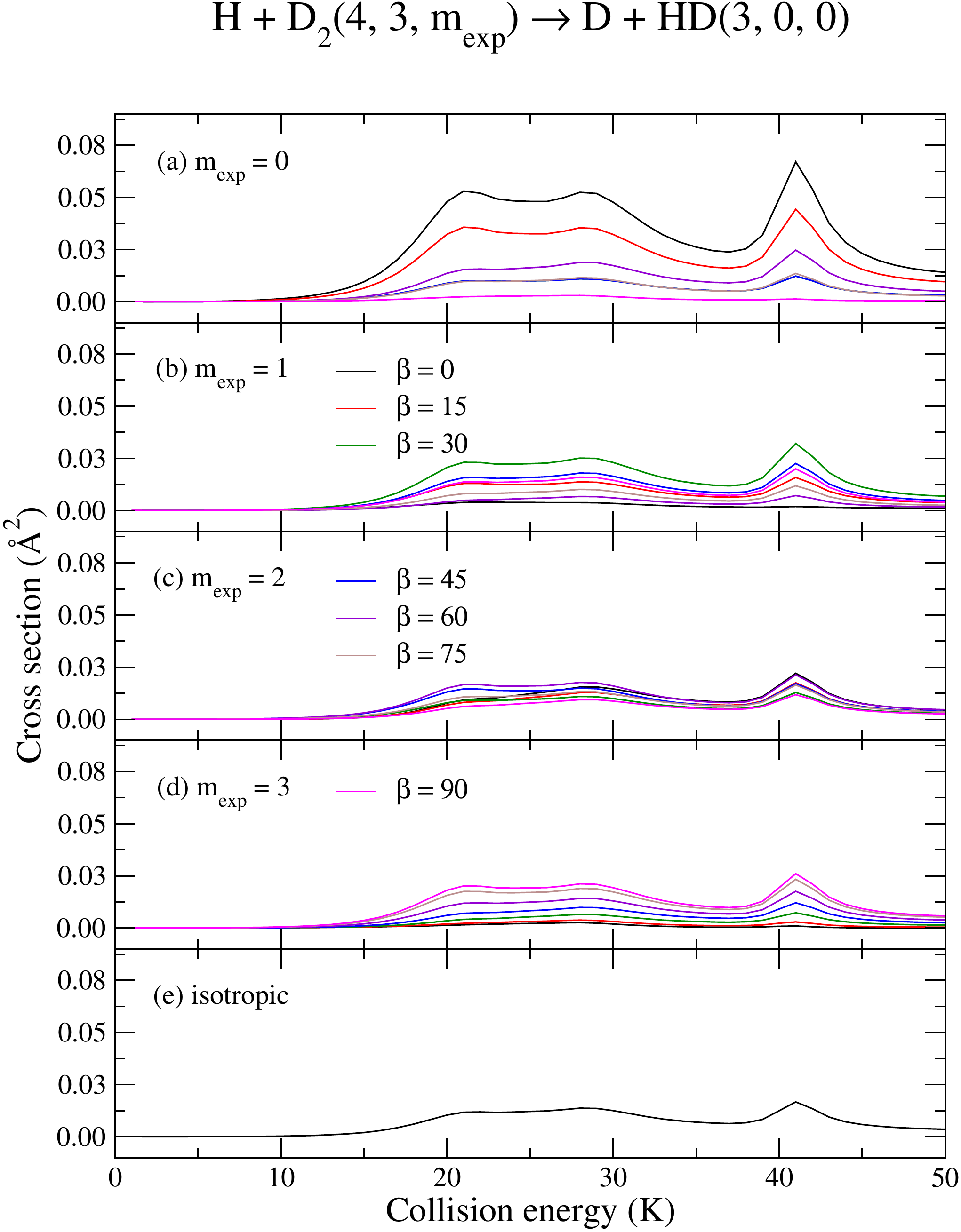}

\caption{\label{fig:ICSvsCollEnergy43-3First} Panels (a)\textendash{}(d): $m_{\mathrm{exp}}$\textendash{}dependent ICS for the H + D$_{2}\left(\upsilon=4,j=3,m_{\mathrm{exp}}=0,1,2,3\right)\protect\longrightarrow$ D + HD$\left(\upsilon^{\prime}=3,j^{\prime}=0,m^{\prime}=0\right)$ chemical reaction as functions of the collision energy, with the same color codes  as in Fig. (\ref{fig:ICSvsCollEnergy44-30Second}).}
\end{figure}

Figure (\ref{fig:ICSvsCollEnergy43-3First}) displays the result for the $j=3$  rotational state of D$_2$ for all relevant values of $m_{\mathrm{exp}}$ and selected values of $\beta$, yielding HD molecules in the $j^{\prime}=0$ rotational state $\left(\upsilon^{\prime}=3\right)$. For this case $J_{\mathrm{max}}=7$ is used and satisfactory numerical convergence is obtained with $J_{\mathrm{max}}=6$.

Results presented in panels (a)\textendash{}(e) of Fig. (\ref{fig:ICSvsCollEnergy43-3First}) show a strong stereodynamical effect for different choices of $m_{\mathrm{exp}}$ and $\beta$ as for the $j=4$ case. It is worthwhile to reiterate that this dynamical feature is entirely driven by the properties of the Wigner coefficients, $d_{mm_{\textrm{exp}}}^{j}\left(\beta\right)$, regardless of other scattering characteristics under consideration, such as collision energy. Thus, at the main resonant peak ($\approx$ 41 K), reactivity is subjected to an enhancement of about a factor of 6 for a $m_{\mathrm{exp}}=0$, $\beta=0$ preparation and a complete suppression for $m_{\mathrm{exp}}=3$, $\beta=0$. Likewise, reactivity may be effectively suppressed by a $m_{\mathrm{exp}}=0$, $\beta=90^{\circ}$ preparation while enhanced at $m_{\mathrm{exp}}=3$, $\beta=90^{\circ}$. The main partial wave contributions for the broad double--peaked resonant structure are $\ell=$ 2--5 whereas the peak at about 41 K is mostly composed of $\ell=$ 4 and $\ell=$ 6.

The resonance structures depicted in Figs. (\ref{fig:ICSvsCollEnergy44-30First}\textendash{}\ref{fig:ICSvsCollEnergy43-3First}) span a somewhat larger range of collision energies, in which a non\textendash{}negligible number of partial waves contribute. For the SARP experiments, where four\textendash{}vector correlation measurements are used to extract scattering properties, lower collision energies (and a limited number of partial waves) are desirable~\cite{perreault2017}. Below we present  results for D$_{2}$ prepared in the $j=2$ rotational state ($J_{\mathrm{max}}=4$) that depict several resonances in the 5\textendash{}10 K regime accessible in a SARP experiment.

\begin{figure}
\includegraphics[scale=0.43]{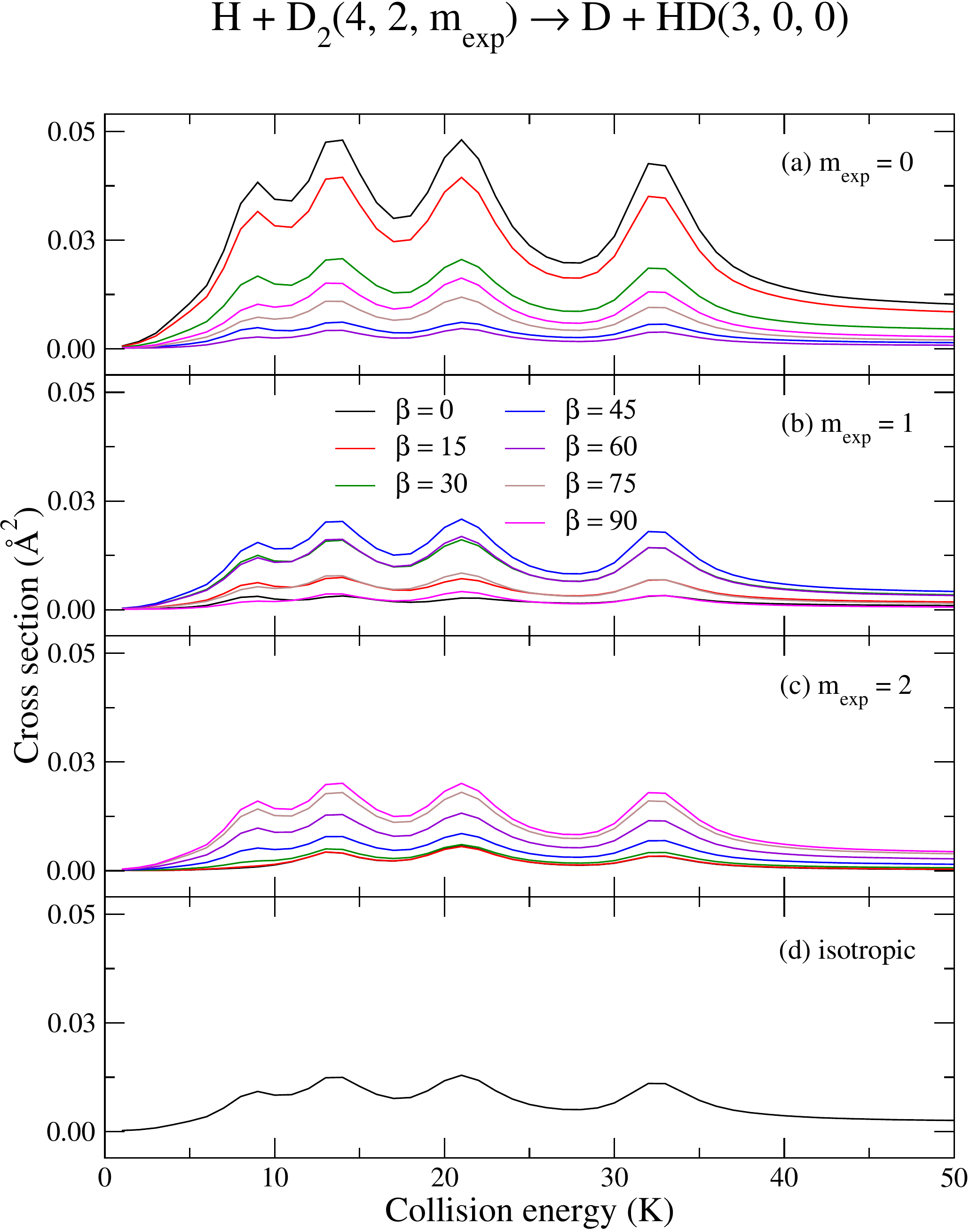}

\caption{\label{fig:ICSvsCollEnergy42First}Panels (a)\textendash{}(c): $m_{\mathrm{exp}}$\textendash{}dependent
ICS for the H + D$_{2}\left(\upsilon=4,j=2,m_{\mathrm{exp}}=0,1,2\right)\protect\longrightarrow$
D + HD$\left(\upsilon^{\prime}=3,j^{\prime}=0,m^{\prime}=0\right)$
chemical reaction as function of the collision energy, with the same color codes as in Fig. (\ref{fig:ICSvsCollEnergy44-30Second}).}
\end{figure}

\begin{figure}
\includegraphics[scale=0.43]{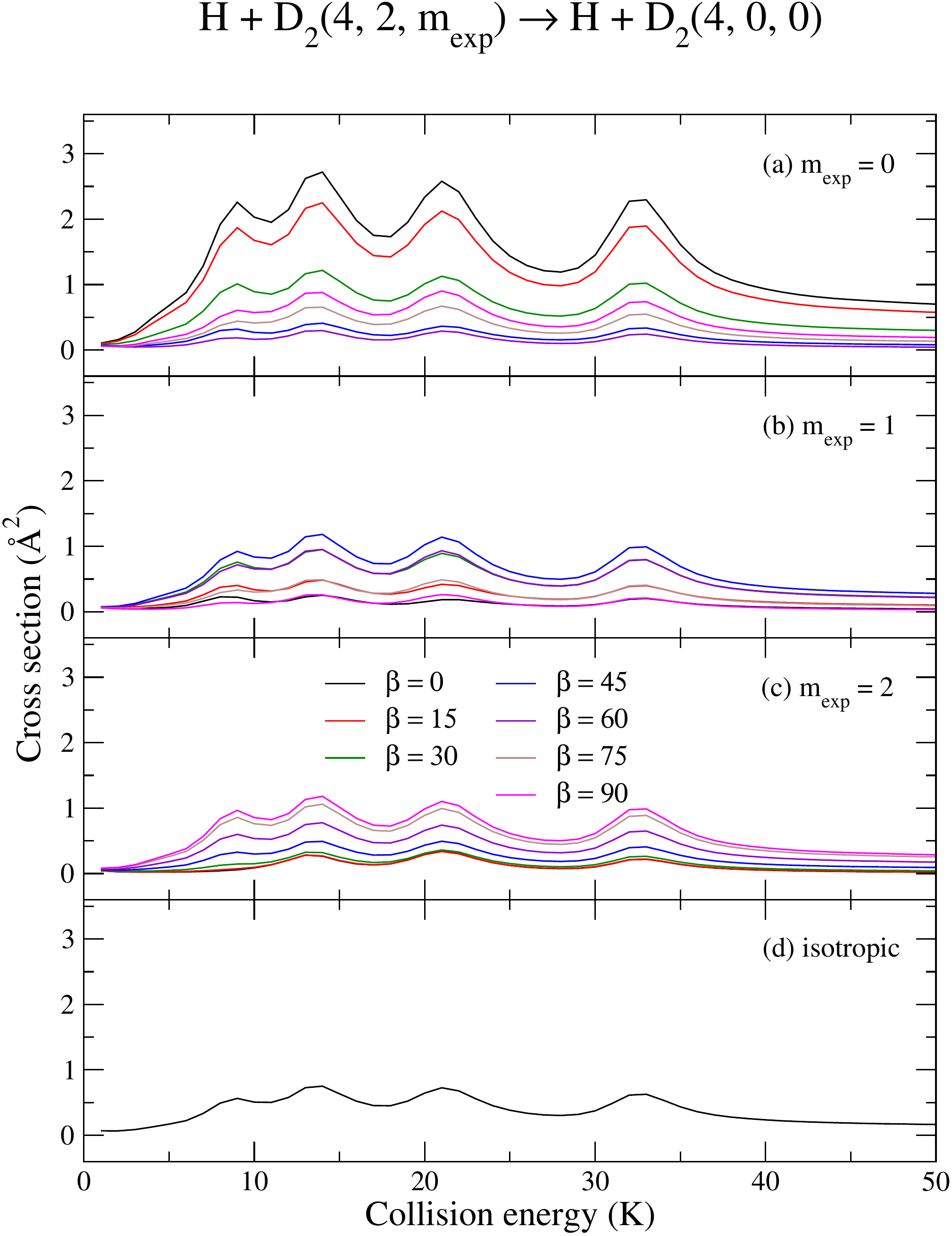}

\caption{\label{fig:ICSvsCollEnergy42Second}Panels (a)\textendash{}(c): $m_{\mathrm{exp}}$\textendash{}dependent
ICS for the H + D$_{2}\left(\upsilon=4,j=2,m_{\mathrm{exp}}=0,1,2\right)\protect\longrightarrow$
H + D$_{2}\left(\upsilon^{\prime}=4,j^{\prime}=0,m^{\prime}=0\right)$ inelastic collision, with the same color codes as in Fig. (\ref{fig:ICSvsCollEnergy44-30Second}).}
\end{figure}

Panels (a)\textendash{}(d) of Fig. (\ref{fig:ICSvsCollEnergy42First}) show the energy dependence of the ICS for the H + D$_{2}\left(\upsilon=4,j=2,m_{\mathrm{exp}}=0,1,2\right)$ $\longrightarrow$ D + HD$\left(\upsilon^{\prime}=3,j^{\prime}=0,m^{\prime}=0\right)$ reaction, whereas those panels of Fig. (\ref{fig:ICSvsCollEnergy42Second}) present the ICS associated to the collision\textendash{}induced de\textendash{}excitation of D$_{2}$, from $j=2$ to $j^{\prime}=0\,\left(\upsilon=\upsilon^{\prime}=4\right)$. The first resonance occurs at about 9 K whose composition is dominated by $\ell=1$ and $\ell=3$ with a minor background contribution from $\ell=2$. At about 13 K a second (and more intense) peak is predicted to be formed from $\ell=$ 0, 2 and 4, followed by another occurrence at 20 K $\left(\ell=1,3,5\right)$ and 32 K $\left(\ell=4,6\right)$. At energies as low as 1 K, most of the contribution is expected from $\ell=0$ and $\ell=2$ followed by a minor, yet significant, contribution from $\ell=1$. In terms of an actual experimental realization at $E_{\mathrm{coll}}\approx5$\textendash{}10 K, therefore, $j=2$ is likely to offer the best conditions when considering H + D$_{2}$ collisions.

Cross sections for the inelastic process \textendash{} panels (a)\textendash{}(d) of Fig. (\ref{fig:ICSvsCollEnergy42Second})  \textendash{} are presented to illustrate how the exact same machinery described so far applies in these cases as well. The inelastic cross sections are about a factor of 60 larger than their reactive counterparts, due in part, to the direct and completely barrierless pathway for the inelastic process. Panel (c) of both figures unveil another intriguing aspect, where the first resonant feature ($\approx$ 9 K) is suppressed entirely for a $m_{\mathrm{exp}}=2$, $\beta\approx0$ preparation. As before, from Eq. (\ref{eq:DiatomRotStateExpansion}), a state $\left|\alpha\upsilon j;m_{\mathrm{exp}}=2\right\rangle =\left|\alpha\upsilon j;m=2\right\rangle $ is expected for a parallel orientation $\left(\beta=0\right)$, whereas at $\beta=15^{\circ}$ would yield
\begin{equation}
\begin{aligned}
\left|\alpha\upsilon j;m_{\mathrm{exp}}=2\right\rangle & =0.0003\left|\alpha\upsilon j;m=-2\right\rangle \\
 & -0.004\left|\alpha\upsilon j;m=-1\right\rangle \\
 & +0.041\left|\alpha\upsilon j;m=0\right\rangle \\
 & -0.254\left|\alpha\upsilon j;m=+1\right\rangle \\
 & +0.966\left|\alpha\upsilon j;m=+2\right\rangle
\end{aligned}
\end{equation}
which is strongly dominated by the $m=+2$ component. However, at about 9 K, for $\ell=1,3$ and $j=2$, the most relevant contributions arise from $m=0$ and $m=\pm1$ for $J=$ 1\textendash{}4, whereas for $J>2$,  contributions from $m=\pm2$  become more relevant above 10 K. Hence, all resonances above 10 K predicted for a D$_2$ state prepared with $j=2,m_{\mathrm{exp}}=2,\beta\approx0$ shall survive mostly due to their $m=2$ character. Therefore, these calculations suggests that controlling the relative orientations of reactants can be also exploited in terms of addressing specific resonances, systematically, due to their $m$\textendash{}, $\ell$\textendash{} and $J$\textendash{}composition, even if they happen to occur in a band composed of overlapping structures. Moreover, an inspection of the energy dependence of $\beta$\textendash{}dependent ICS presented up to this point reveals an intricate competition among different $\beta$ values. As the energy varies, the actual $m\rightarrow m^{\prime}$ branching ratio in the S\textendash{}matrix is likely to vary whereas, at same time, $\beta$ controls the weight of various $m$\textendash{}terms in a given $j,m_{\mathrm{exp}}$ preparation. Morita and Balakrishnan noted a similar observation~\cite{morita2020b}, in terms of an interference effect due to cross terms in the square of the scattering amplitude, in their analysis of stereodynamical effects in rotationally inelastic collisions of HD and He.

\begin{figure}
\includegraphics[scale=0.43]{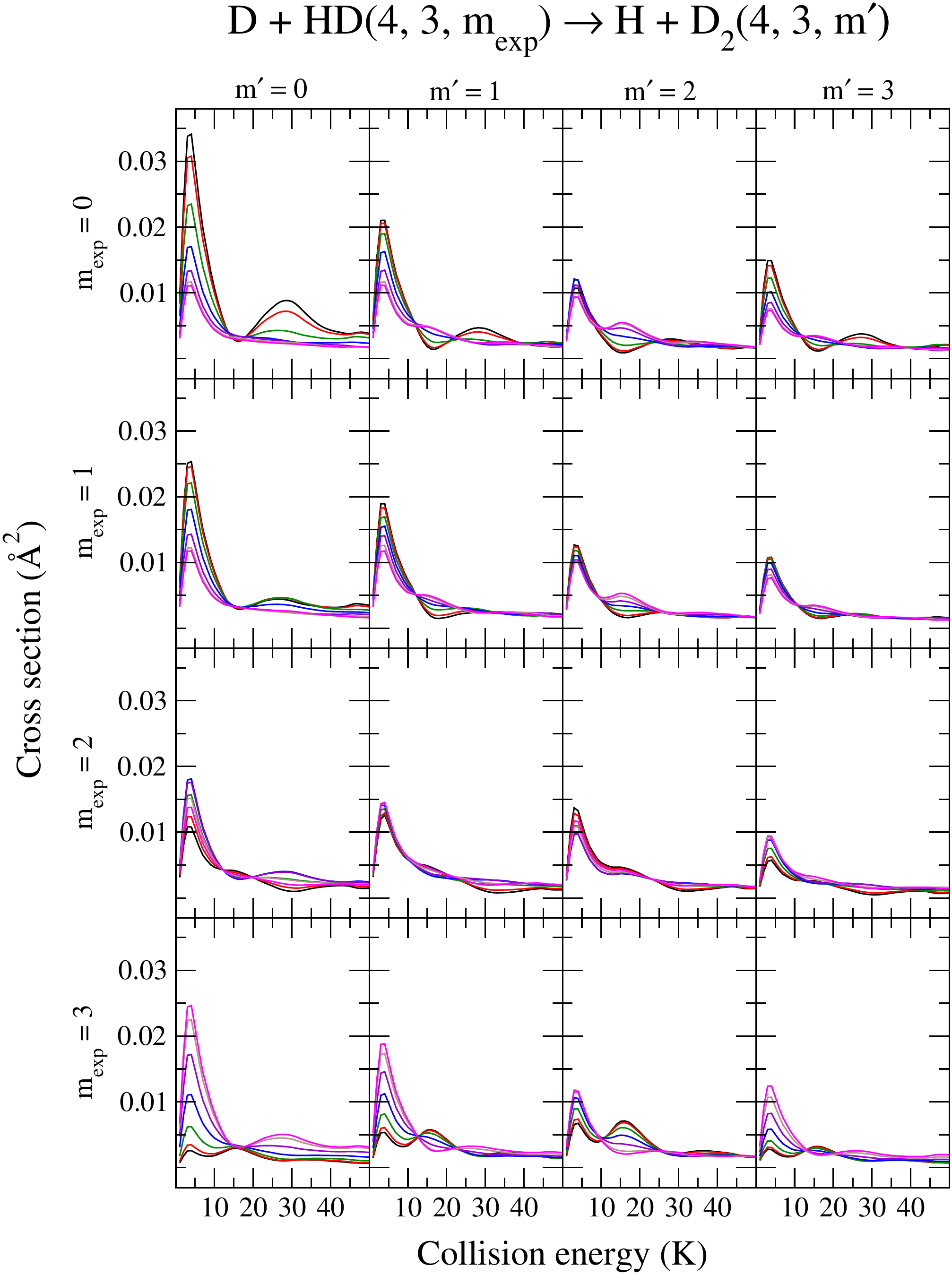}

\caption{\label{fig:ICSvsCollEnergy43-43} Each panel corresponds to a $\left(m_{\mathrm{exp}},m^{\prime}\right)$\textendash{}dependent ICS for the D + HD$\left(\upsilon=4,j=3,m_{\mathrm{exp}}=0,1,2,3\right)\protect\longrightarrow$ H + D$_{2}\left(\upsilon^{\prime}=4,j^{\prime}=3,m^{\prime}=0,1,2,3\right)$ chemical reaction as functions of the collision energy, with the same color codes  as in Fig. (\ref{fig:ICSvsCollEnergy44-30Second}).}
\end{figure}

We wrap up the discussion of ICSs by presenting results for the reverse reaction, D + HD $\to$ H + D$_2$, Fig. (\ref{fig:ICSvsCollEnergy43-43}), in which HD is prepared in $j=3$ ($J_{\mathrm{max}}=7$) with $m_{\mathrm{exp}}\leq3$ for all $\beta$ values discussed previously. We limit our analysis to $j^{\prime}=3$ of D$_2$ to inspect all 16 possible $\left(m_{\mathrm{exp}},m^{\prime}\right)$\textendash{}cases, where $m^{\prime}\leq3$ (negative values are not shown for the sake of simplicity).

Many aspects of the results highlighted for  H + D$_{2}$ are also observed here, and some are highlighted below:

(i) In panel $m_{\mathrm{exp}}=0,m^{\prime}=0$, the maximum enhancement of reactivity is observed for the $\beta=0$ preparation and cross sections are suppressed for higher $\beta$ values.

(ii) In panel $m_{\mathrm{exp}}=3,m^{\prime}=0$: enhancement is found for higher $\beta$ values.

(iii) In panel $m_{\mathrm{exp}}=0-3,m^{\prime}=2$: a unique resonant feature at about 15 K is predicted to emerge from the $m^{\prime}=2$ case, whereas smaller traces of it are also expected for other $m^{\prime}>0$ cases.

(iv) In panels $m_{\mathrm{exp}}=0,m^{\prime}=1-3$: selective suppression of a specific resonance occurs with increase in $m^{\prime}$ at $E_{\mathrm{coll}}\approx$ 20\textendash{}40 K as $\beta$ varies from 0 to 90$^\circ$.

\begin{figure}
\includegraphics[scale=0.4]{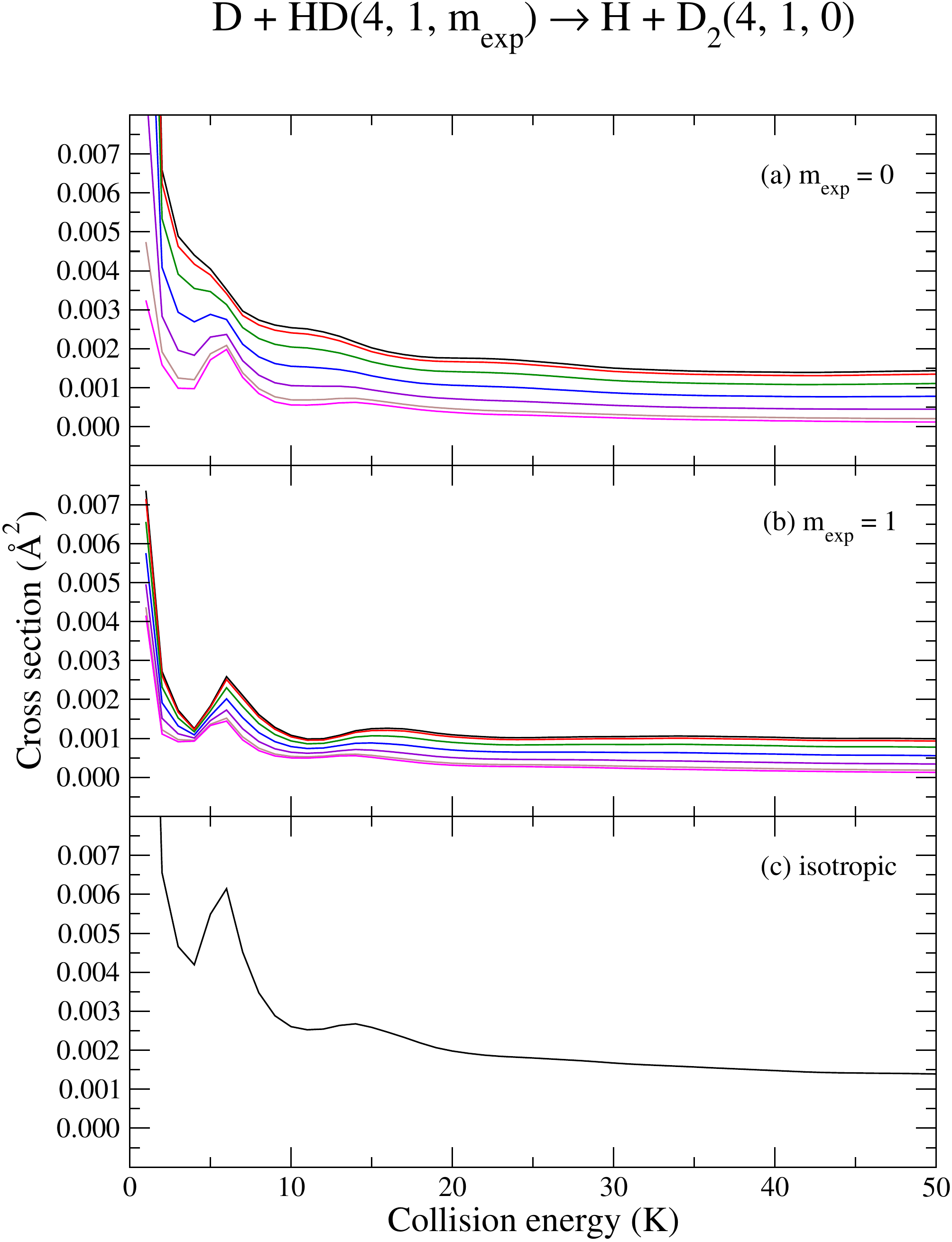}

\caption{\label{fig:ICSvsCollEnergy41-41} Panels (a) and (b): $m_{\mathrm{exp}}$\textendash{}dependent ICS for the D + HD$\left(\upsilon=4,j=1,m_{\mathrm{exp}}\right)\protect\longrightarrow$ H + D$_{2}\left(\upsilon^{\prime}=4,j^{\prime}=1,m^{\prime}=0\right)$ chemical reaction as functions of the collision energy, with the same color codes  as in Fig. (\ref{fig:ICSvsCollEnergy44-30Second}).}
\end{figure}

The richness, in terms of resonant and stereodynamical features, manifested in the ICS for a collision of D with HD, with HD prepared in $j=3,m_{\mathrm{exp}}$, offers various handles for stereodynamic control of this reaction, in particular, for those experiments targeting collision energies near 1 K. Similarly, the $j=1$ rotational levels from both HD and D$_{2}$ are also predicted to yield higher reactivity due to a dominant $\ell=1$ and $\ell=2$ contribution near $E_{\mathrm{coll}}\approx 1$ K. This is illustrated in Fig. (\ref{fig:ICSvsCollEnergy41-41}) for the D + HD$\left(\upsilon=4,j=1,m_{\mathrm{exp}}\right)\longrightarrow$ H + D$_{2}\left(\upsilon^{\prime}=4,j^{\prime}=1,m^{\prime}=0\right)$ reaction ($J_{\mathrm{max}}=4$). A second resonant feature, whose properties are modulated by $m=\pm1$ and a $\ell=3$ partial wave, occurs at about 6 K. This resonance is also very sensitive to a $m_{\mathrm{exp}},\beta=0$ preparation (comparing black curves from panels (a) and (b)). Due to such a marked characteristic and presence of an isolated partial wave occurrence, a collision of D + HD$\left(\upsilon=4,j=1,m_{\mathrm{exp}}\right)$ at about 6 K, with $m_{\mathrm{exp}}=0,\beta=0$ and $m_{\mathrm{exp}}=1,\beta=0$ preparations, appears to be a good candidate for an experimental study. Panel (c) of Fig. (\ref{fig:ICSvsCollEnergy41-41}) depicts the isotropic cross section, where a smaller resonance is also expected to occur at about 14 K ($\ell=4$).

\begin{figure}
\includegraphics[scale=0.4]{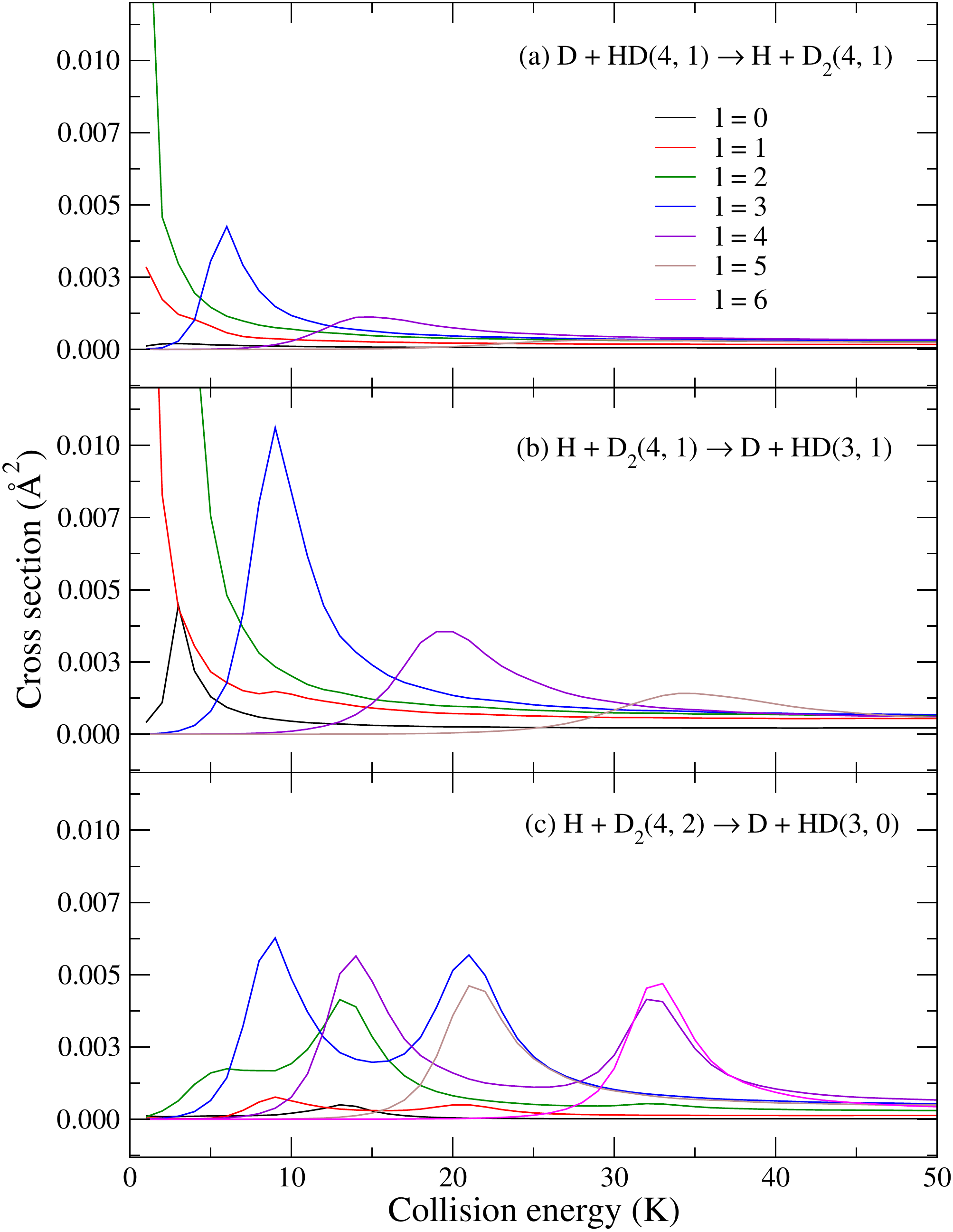}

\caption{\label{fig:ICSPartialWaves} Panels (a)\textendash{}(c): $\ell$\textendash{}dependent isotropic ICS as functions of the collision energy with each panel for one of the displayed reactions;  $\ell=0$ (black curve), $\ell=1$ (red curve), $\ell=2$ (green curve), $\ell=3$ (blue curve), $\ell=4$ (violet curve), $\ell=5$ (brown curve) and $\ell=6$ (magenta curve).}
\end{figure}

In Fig. (\ref{fig:ICSPartialWaves}) we present the partial wave composition for the isotropic ICS (summed over  $J=0$\textendash{}4) for the D + HD$\left(\upsilon=4,j=1\right)\to$ H + D$_2\left(\upsilon^{\prime},j^{\prime}\right)$ and H + D$_2\left(\upsilon=4,j=1,2\right)\to$  D + HD$\left(\upsilon^{\prime},j^{\prime}\right)$ reactions. Panel (a) depicts results for $\upsilon^{\prime}=4$ and $j^{\prime}=j=1$; panel (b) for $\upsilon^{\prime}=3$ and $j^{\prime}=j=1$; and, panel (c) for $\upsilon^{\prime}=3$ and $j^{\prime}=0$. The different panels show peaks associated with specific partial waves where $\ell=2$ dominates near 1 K followed by smaller contributions of $\ell=1$. Therefore, $j=1$ is predicted as an ideal case for experimental investigations of both D + HD and H + D$_{2}$ reactions near 1 K whereas $j=2$ is preferred for measurements from 5 K ($\ell=2$ and $\ell=3$) to about 35 K ($\ell=4$ and $\ell=6$). We would like to stress that the marked resonance profiles discussed in this work are a manifestation of a given entrance channel $n$. Thus, other choices of exit channels, $n^{\prime}$, shall present a similar energy\textendash{}dependent behavior and no compromise has been made by choosing a specific $\upsilon^{\prime}$ or $j^{\prime}$ product state.

\subsection{Differential cross sections}

Now we consider how the relative orientation of reactants influences the deflection of products towards a given scattering angle $\theta$. To this end we shall focus on the $m_{\mathrm{exp}}=0$ preparation as it has been shown to provide the largest enhancements, upon proper choices of $\beta$. As energy\textendash dependent measurements of scattering processes are inherently more complicated, we shall also fix the collision energies to 1 K, 5 K and 10 K.

\begin{figure}
\includegraphics[scale=0.43]{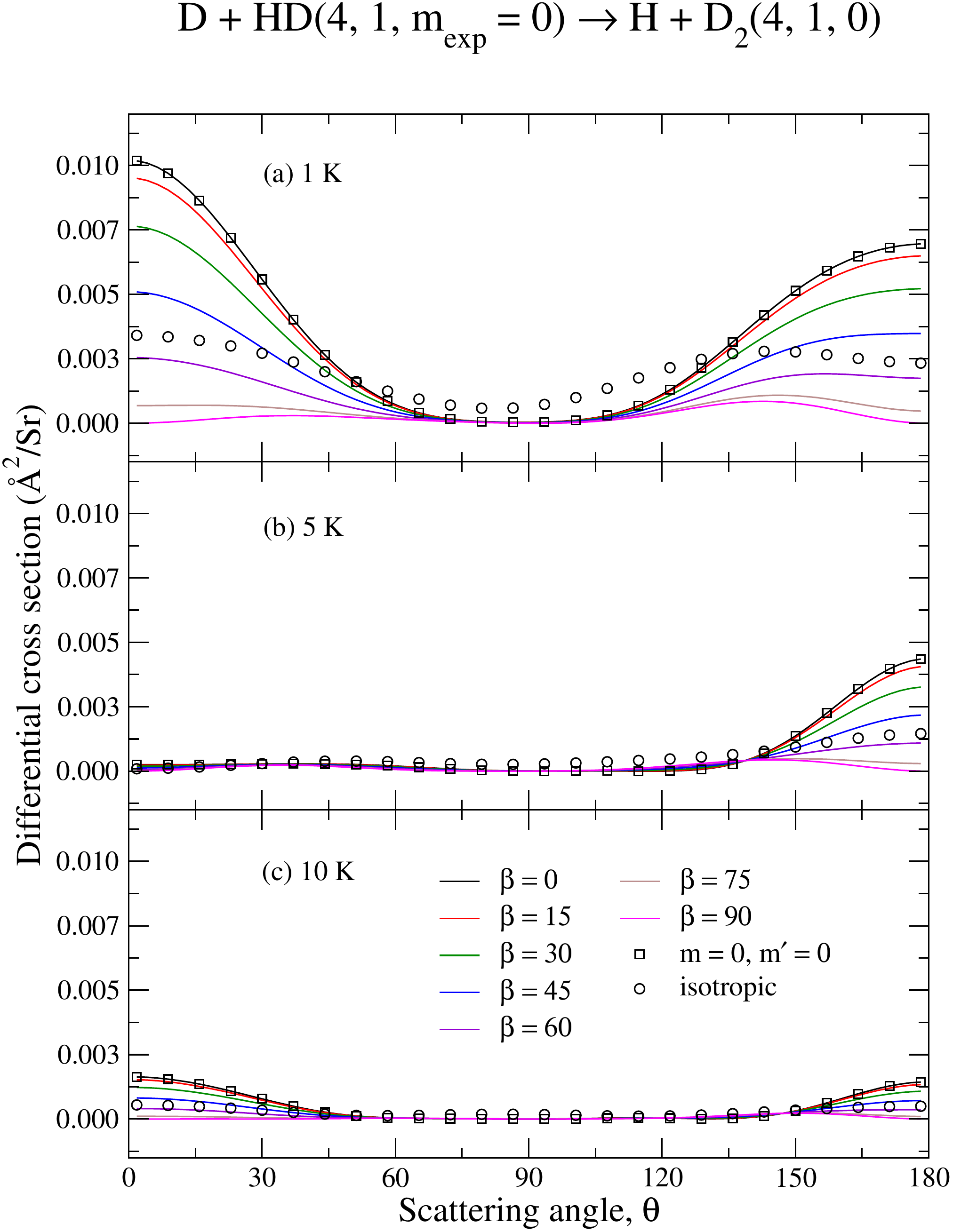}

\caption{\label{fig:HD-DCS-First}Panels (a)\textendash{}(c): DCS for the D + HD $\left(\upsilon=4,j=1,m_{\mathrm{exp}}=0\right)\protect\longrightarrow $H + D$_{2}\left(\upsilon^{\prime}=4,j^{\prime}=1,m^{\prime}=0\right)$ chemical reaction as functions of the scattering angle $\theta$, for collision energies of 1 K, 5K and 10 K, respectively, where the color codes are the same as in Fig. (\ref{fig:ICSvsCollEnergy44-30Second}).}
\end{figure}

\begin{figure}
\includegraphics[scale=0.43]{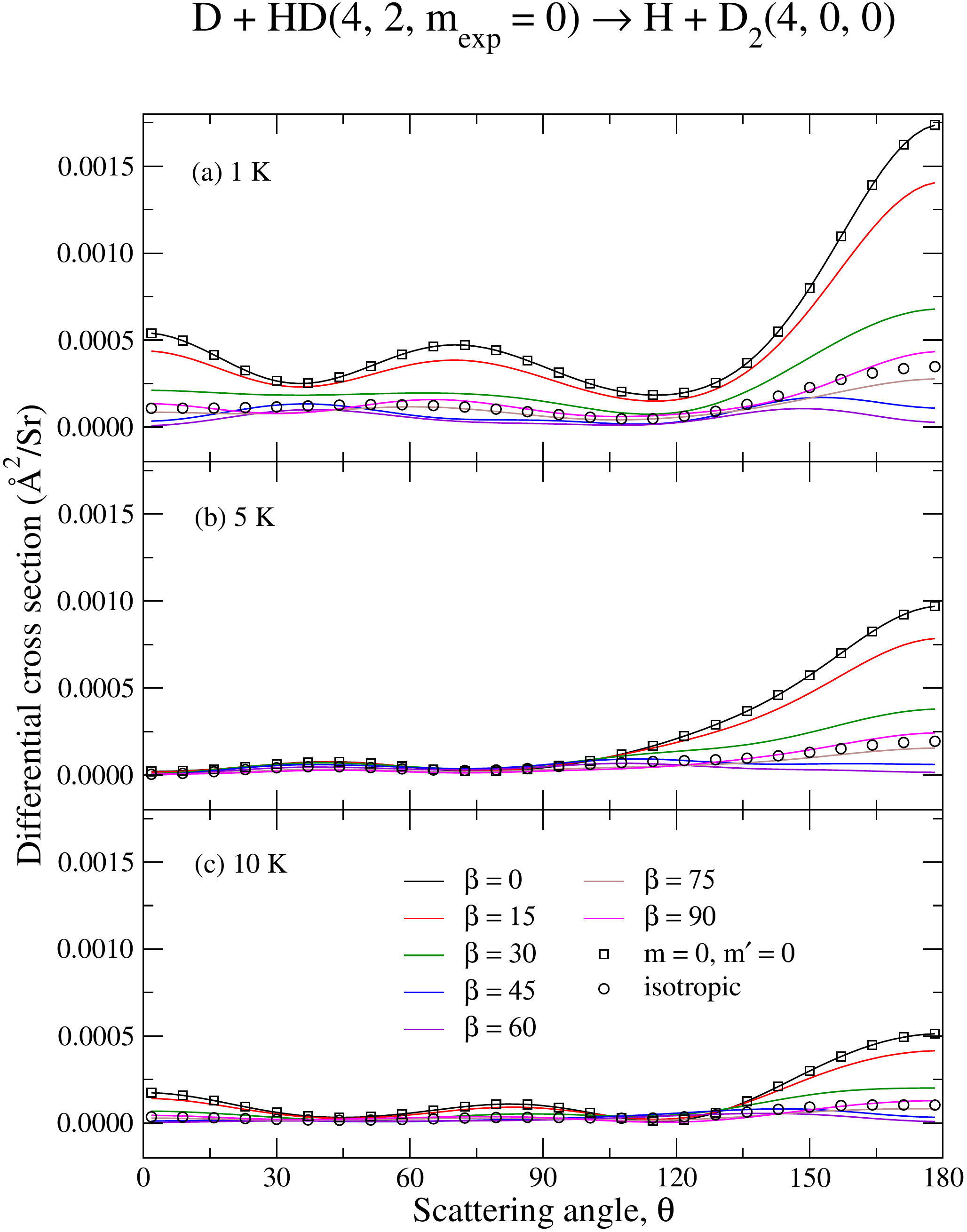}

\caption{\label{fig:HD-DCS-Second}Panels (a)\textendash{}(c): the same as Fig. (\ref{fig:HD-DCS-First}) but for $j=2$ and $j^{\prime}=0$.}
\end{figure}

Panels (a)\textendash{}(c) from Fig. (\ref{fig:HD-DCS-First}) present the $\theta$\textendash{}dependency of the DCS for the D + HD$\left(\upsilon=4,j=1,m_{\mathrm{exp}}=0\right)\to$ H + D$_{2}\left(\upsilon^{\prime}=4,j^{\prime}=1,m^{\prime}=0\right)$ reaction, whereas panels (a)\textendash{}(c) of Fig. (\ref{fig:HD-DCS-Second}) present a similar result for the $j=2$ and $j^{\prime}=0$ case. As before, for comparison purposes, circles are used for the averaged isotropic DCS and squares are used for the pure $m=0,m^{\prime}=0$ component (summed over $J=$ 0\textendash{}4). A comparison between the isotropic case (circles) and the  $m$\textendash{}to\textendash{}$m^{\prime}$ component (squares) unveils once again a similar mechanism discussed above in terms of ICS: \textit{i.e.}, the collision\textendash{}induced branching ratio associated to lower $m$\textendash{}values, mainly $m=0$, is likely to contribute the most but is often suppressed due to the typical average over all $2j+1$ $m$\textendash{}terms. However, the dominant contribution from $m=0$ is unlocked by choosing $\beta=0$ (black curves). As the molecular alignment angle changes from $\beta= 0\rightarrow{}90^{\circ}$, the relative contributions of $m=0$ become smaller in $\left|\alpha\upsilon j;m_{\mathrm{exp}}=0\right\rangle $ yielding, therefore, an overall smaller cross section.

The preference for (backward) forward scattering $\left(\theta\rightarrow0,\theta\rightarrow\pi\right)$ is seen to be mostly captured by the $m=0,m^{\prime}=0$ component and this preference is either magnified $\left(\beta\rightarrow0\right)$ or suppressed $\left(\beta\rightarrow90^{\circ}\right)$ by a given choice of $\beta$. By comparing results depicted by circles and squares again, one notices that the small isotropic DCS value at $\theta=90^{\circ}$ includes a negligible, if any, contribution from the $m=0,m^{\prime}=0$ component. Thus, the DCS is the smallest and almost independent of the scattering angle for $\beta=90^{\circ}$.

\begin{figure}
\includegraphics[scale=0.4]{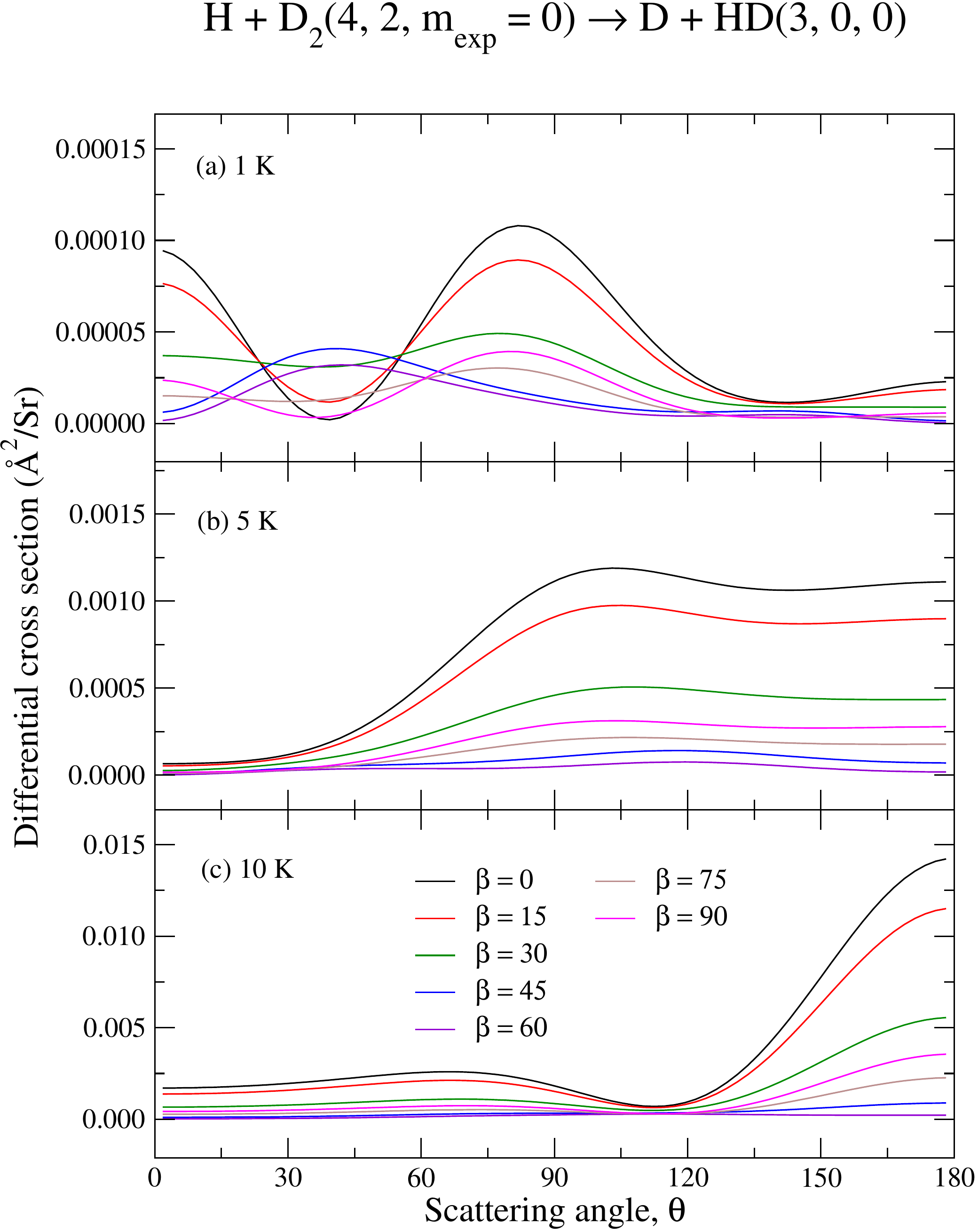}

\caption{\label{fig:D2-DCS}The same as Fig. (\ref{fig:HD-DCS-First}) but for the H + D$_{2}\left(\upsilon=4,j=2,m_{\mathrm{exp}}=0\right)\protect\longrightarrow$ D + HD$\left(\upsilon^{\prime}=3,j^{\prime}=0,m^{\prime}=0\right)$ reaction.}
\end{figure}

For the sake of completeness, Fig. (\ref{fig:D2-DCS}) illustrates a typical DCS for the H + D$_{2}\left(\upsilon=4,j=2,m_{\mathrm{exp}}=0\right)\protect\longrightarrow$ D + HD$\left(\upsilon^{\prime}=3,j^{\prime}=0,m^{\prime}=0\right)$ reaction. In this case, the DCS at 1 K is peaked near 90$^{\circ}$ but fairly negligible overall. As the collision energy increase to 5 K and 10 K, the scattering becomes preferentially backward with maximum reactivity attained at 10 K, in which case, from Fig. (\ref{fig:ICSPartialWaves}), a $\ell=3$ character is predicted. At all three energies $\beta\le 15^{\circ}$ favors the reaction.

\section{Conclusion and prospects \label{sec:summary}}

In this work we report extensive quantum dynamics calculations of the H + D$_2$ and D + HD chemical reactions with an aim of
controlling the reaction outcome through stereodynamic preparation of the molecular rotational state. Our approach extends the formalism presented in recent experimental studies of Perreault \textit{et al.} \cite{perreault2017,perreault2018,perreault2019,sarp_hed2} applied to rotational quenching of HD by H$_2$, D$_2$ and He and D$_2$ by He to state\textendash{}to\textendash{}state chemistry for the first time. We report vast control through stereodynamic preparation of the molecular state, essentially allowing switching the reaction on or off through appropriate alignment of the molecular bond axis relative to the direction of the incident relative velocity vector.

The computations are performed using a state\textendash{}of\textendash{}the\textendash{}art time\textendash{}independent quantum reactive scattering formalism in hyperspherical coordinates at a moderate range of collision energies, 1\textendash{}50 K. Both diatomic molecules were prepared in excited rovibrational states, $\upsilon=4,j=$ 1\textendash{}4, and several resonant features have been predicted to occur. Intensities of these resonances were found to be strongly sensitive to the stereodynamic preparation of the molecular rotational states. Several partial wave resonances were identified in this study with $\ell=$ 0\textendash{}2 dominating at lower collision energies (below 10 K) and $\ell=$ 4\textendash{}6 at higher energies. To our knowledge, resonances for the initial states considered here $\left(\upsilon=4,j>0,E_{\mathrm{coll}}=0-50\,\mathrm{K}\right)$ are predicted for the first time and deserve additional investigations in order to characterize properties such as lifetime, quantum numbers etc. Our findings are significant in light of recent experimental progress in preparing vibrationally excited H$_2$ in the $v=7$ vibrational level through the SARP techniques allowing the prospects of studying chemical reactions involving highly vibrationally excited  molecules that are also prepared in a stereodynamic state \cite{perreault2020}.

Our results reveal the following aspects which may or may not be universal to the specific system investigated here:

(i) The isotropic reactive cross section, associated to freely rotating reactants, is composed of contributions from smaller $m$\textendash{}values, in particular, $m=0\left(m^{\prime}\approx0\right)$. However, as one averages over all $2j+1$ initial $m$ states, these higher contributions are often washed out.

(ii) By imposing a certain spatial polarization on the rotational state $\mathbf{j}$ of reactants, a particular $m$\textendash{ contribution (denoted by $m_{\mathrm{exp}}$ in this work) can be filtered by rotating its internuclear axis.

(iii) The stereodynamically prepared state becomes a pure state as $\beta\rightarrow0$, \textit{i.e.}, $\pm m_{\mathrm{exp}}\rightarrow \pm m$ as $\beta\rightarrow0$, where $\beta$ is the alignment angle of the molecular bond axis relative to the incident velocity vector, \textit{i.e.} orientation of the SARP polarization lasers relative to the molecular beam in the experiments of Perreault \textit{et al.}

(iv) The prepared state $m_{\mathrm{exp}}$ becomes a broad mixture of all $m$\textendash{}terms as $\beta \rightarrow 90^{\circ}$.

(v) Due to (iii), the $m=0$ contribution (largest for the present system) are accessible by a $m_{\mathrm{exp}}=0,\beta=0$ preparation, which generally enhances the collision\textendash{}induced branching ratio of interest. This is equivalent to the H--SARP preparation in the SARP experiments.

(vi) Likewise, due to (iv), in order to enhance unfavored branching ratios, often associated to higher $m$ (and $m^{\prime}$) values, a preparation combining higher $m_{\mathrm{exp}}$ with $\beta \rightarrow 90^{\circ}$ can be used.

(vii) A lower value of $m_{\mathrm{exp}}$ combined with a higher value of $\beta$ (and vice\textendash{}versa) mostly suppresses the reaction.

Possibly, without loss of generality, points (i)\textendash{}(vii) can be viewed as a set of propensity rules that might be exploited
to effectively control any cold chemical reaction, as well as inelastic collisions.

At collision energies above a few Kelvin a diverse array of stereodynamical effects are predicted to occur due to the participation of various partial waves and different $m$\textendash components of the rotationally excited states that allow additional control of the reaction outcome. Our results show that both the collision energy and the alignment angle $\beta$ have a strong effect on the $m\rightarrow m^{\prime}$ branching ratios and the $m$\textendash composition of a given $m_{\mathrm{exp}}$ state, respectively. Likewise, we have found strong evidences suggesting that particular resonances can be selectively controlled, by means of orienting the reactant appropriately, due to their specific partial wave, $m$ and $m^{\prime}$ compositions.

Such a degree of control of a chemical reaction involving neutral systems, in a completely field\textendash{}free environment, requires only ro--vibratational excitation, which is relatively easy to realize experimentally and it is also a universal property in contrast to, for instance, the need of working with electrically polar molecules. In particular, two\textendash{}body
losses in ultracold traps $\left(E_{\mathrm{coll}}\ll1\,\mathrm{K}\right)$, due to chemical reactions and inelastic collisions, have been a long standing problem, mostly circumvent by means of an elegant use of external fields that prevents the colliding partners to approach one another \cite{avdeenkov2006,wang2015,martinez2017}. Besides their effectiveness, most of these methods are not general and a given molecular system of interest must comply with specific electric properties. In contrast, we have demonstrated here how appropriate choices of $m_{\mathrm{exp}}$ and $\beta$ \textendash{} namely, lower (higher) $m_{\mathrm{exp}}$ and higher (lower) $\beta$ \textendash{} is mostly effective in ``turning off'' the chemical reaction in a very general fashion.

Finally, in terms of actual experimentation using H + D$_{2}$ and D + HD, our calculations suggest that choices of either $j=2$ or $j=1\left(\upsilon=4\right)$ are the best compromises in order to explore resonances at sufficiently low energies, where only few partial waves contribute. In particular, near 1 K, preparing either D + HD or H + D$_{2}$ with $j=1$ is a promising choice. However, if the presence of higher partial waves are less of a problem experimentally, and higher collisional energies are of interest, then $j=2$ and $j=3\left(\upsilon=4\right)$ are better candidates as they are predicted to possess a series of overlapping resonant structures over a broader range of collision energies.

\section*{Conflicts of interest}
There are no conflicts to declare.


\begin{acknowledgments}
N.B. acknowledges support from NSF grant No. PHY-1806334 \& PHY-2110227 as well as ARO MURI grant No. W911NF-19-1-0283. This work used the Extreme Science and Engineering Discovery Environment (XSEDE), which is supported by National Science Foundation grant number ACI-1548562. Specifically, it used the Bridges-2 system, which is supported by NSF award number PHY-200034 (N.B.) at the Pittsburgh Supercomputing Center (PSC). B.K.K. acknowledges that part of this work was done under the auspices of the US Department of Energy under Project No. 20170221ER of the Laboratory Directed Research and Development Program at Los Alamos National Laboratory. This work used resources provided by the Los Alamos National Laboratory Institutional Computing Program. Los Alamos National Laboratory is operated by Triad National Security, LLC, for the National Nuclear Security Administration of the U.S. Department of Energy (Contract No. 89233218CNA000001).
\end{acknowledgments}

\section*{Data availability statement}
The data that support the findings of this study are available within the article.

\bibliography{references}

\end{document}